\documentclass[preprint,aps,prc,nofootinbib,showpacs]{revtex4}

\usepackage{graphicx}
\usepackage{graphics}
\usepackage{epsfig}
\usepackage{color}
\usepackage{amssymb}
\usepackage{amsmath}
\usepackage{amsfonts}

\begin{document}
\title{Thermodynamics and critical behavior in the Nambu-Jona-Lasinio model of QCD}
\author{P. Costa}
\email{pcosta@teor.fis.uc.pt}
\author{M. C. Ruivo}
\email{maria@teor.fis.uc.pt}
\affiliation{Departamento de
F\'{\i}sica, Universidade de Coimbra, P-3004-516 Coimbra, Portugal}

\author{C. A. de Sousa}
\email{celia@teor.fis.uc.pt}
\affiliation{Departamento de
F\'{\i}sica, Universidade de Coimbra, P-3004-516 Coimbra, Portugal}
\date{\today}
\begin{abstract}

We investigate the phase diagram of strongly interacting matter as a function of
temperature and baryonic density/chemical potential, within Nambu--Jona-Lasinio  type
models. We perform a systematic study concerning the existence, location, and properties
of a critical end point/tricritical point, both in SU(2) and SU(3) versions of the model.
We verify that, for $m_u=m_d=0$ and up to a critical  strange quark mass, there is a
tricritical point, which becomes a critical end point in a world with realistic values of
the current quark masses.
The properties of physical observables, such as the baryon
number susceptibility and the specific heat, are analyzed in the
vicinity of the critical end point, with special focus on their
critical exponents.
The behavior of mesons in the $T-\mu_B(\rho_B)$ plane is analyzed in connection with
possible signatures of partial and effective restoration of chiral
symmetry.
\end{abstract}
\maketitle


\section{Introduction}

Recently there has been encouraging progress on nonperturbative studies
of the QCD thermodynamics which have stimulated a great deal of theoretical
activity. Phenomenological and microscopic models have been developed along
parallel and complementary lines  allowing to predict a rich phase
structure at finite temperature, $T$, and chemical potential, $\mu_B$
\cite{AlfordRW98-99,RappSSV98,HalaszJSSV,Ratti}. The quark gluon plasm (QGP)
is a longstanding theoretical issue since the discovery of the asymptotic
freedom of QCD \cite{HalaszJSSV,Shuryak2006}. Besides the intrinsic theoretical
interest of this subject, such studies are important because they are directly
applicable to the regime under current experimental investigation at the Brookhaven
National Laboratory (BNL) Relativistic Heavy Ion Collider (RHIC). 
In fact, extensive experimental work has been done with heavy-ion collisions at 
CERN and Brookhaven to explore the $T-\mu_B$ phase diagram and look for 
signatures of the QGP.

Theoretical studies have been accumulating a lot of evidence that there exists a critical
end point (CEP) in the phase diagram of strongly interacting matter. Since Fodor and Katz,
who presented a phase diagram with the CEP within lattice calculations
\cite{Fodor:2004JHEP}, remarkable progress in this area has been made. It is an open
question, whether a critical end point exists on the $T-\mu_B$ plane and, particularly,
how to predict  its location. When only thermal effects are concerned, universal
arguments \cite{Wilczek,RajagoplalW93} and lattice simulations \cite{Ukawa} indicate that
the order of the phase transition depends on the masses and flavors of quarks.

Considering also nonvanishing chemical potentials, a variety of models (see e.g.
\cite{Buballa:NPA1998,Buballa:PLB1999}) predict a second order phase transition point in
the phase diagram. This suggests that the phase diagram exhibits a CEP. At this point the
phase transition is of second order and long wavelength fluctuations appear, leading to
characteristic experimental consequences that can be detected by enhanced critical
fluctuations in heavy-ion collisions \cite{Stephanov:1998PRL,Hatta:2003PRD}. So, the
location of the CEP has become an important topic in effective model studies and lattice
calculations.
In fact, the phase diagram and QCD thermodynamics in general are becoming more
transparent due to the combination of research in several areas: perturbative QCD,
effective models, and lattice calculations.

The possible existence of such a point has recently been emphasized and its universal
critical properties have been discussed by several authors in the context of QCD inspired
models \cite{AchwarzKP99,Buballa:NPA1998,Buballa:PLB1999,PLBcosta}.
This point of the phase diagram is the special focus of the present article.

In a previous work \cite{PLBcosta}, we studied the phase diagram focusing our attention
on the CEP and the physics near it, through the behavior of the baryon number
susceptibility and the specific heat; the study was performed  in the framework of the
SU(3) Nambu--Jona-Lasinio (NJL) model. Here, besides extending  the investigation to other
observables, we make a comparative study of the phase diagram   in the SU(2) and SU(3)
NJL models. Since more information can be taken within the simpler version of the NJL
model, this systematic study is expected to provide a better understanding of the
interesting physics around the CEP/TCP (tricritical point).
Our main goal is to locate the critical end point and confront the results with
universality arguments.
Based on the fact that the CEP is a genuine thermodynamic singularity, being considered a
second order critical point, the order parameter and related observables, like
susceptibilities, can provide relevant signatures for phase transitions.
We notice that susceptibilities in general are related to fluctuations through the
fluctuation dissipation theorem, allowing to observe signals of phase transitions  in
heavy-ion reactions \cite{quat7}. The specific heat $C$, which is related to the
event-by-event temperature fluctuation \cite{quat8}, and mean transverse momentum
fluctuations \cite{quat9} in heavy-ion reactions, is also a quantity of interest in our
calculation. These fluctuations should show a divergent behavior near the CEP.
After equilibration, the dense matter created in relativistic heavy-ion collision
will expand along lines of constant entropy per baryon.

We remark that most of the work done in this area has been performed with non strange
quarks only and, when strange quarks are considered,  mixing between the flavors $u$, $d$,
and $s$ has not been taken into account  \cite{Barducci:1994PRD}. Our SU(3) version of
the NJL model includes a term that incorporates the axial anomaly of QCD, and is
responsible for the mechanism of flavor mixing.
We relate the discontinuity of the order parameter to other discontinuities of physical
quantities such as, for  instance, the entropy. We are particularly interested in
confronting our calculation, in what concerns to the notion of a second order phase
transition due to  nonvanishing  current quark masses, with those of any classical mean
field theory. From lattice calculations it is well known that  the strange quark mass
plays a decisive role in the location of the CEP.

On the other hand, information on the nature of excitations and the strength of their
interaction in the QGP would be crucial in the experimental search. Also in this context
it is relevant to confront first-principle based approaches  with the results
of phenomenological models like the NJL model.

We organize the work in four main steps. First, after the presentation of the model
formalism (Sec. II), we discuss the behavior of the equations of state and analyze the
chiral phase transition (Sec. III). The well known universality hypothesis of phase
transitions will be considered. Second, we study  the behavior of relevant physical
quantities in the $T-\mu_B$ plane (Sec. IV). Third, we analyze the phase diagrams in
the $T-\mu_B$ plane looking for the location of the critical end  point and the behavior
of susceptibilities (Sec. V). Finally, we discuss signs of \textit{partial} and
\textit{effective} restoration of chiral symmetry (Sec. VI), looking for the convergence
of chiral partners. We conclude in Sec. VII with a brief summary of our results.


\section{Formulation of the model}\label{model}

The Lagrangian of the SU(3) NJL model
\cite{njl,{Kunihiro:PR1994},Rehberg:1995PRC} is given by
\begin{eqnarray} \label{lagr}
{\mathcal L} &=& \bar{q} \left( i \partial \cdot \gamma - \hat{m} \right) q
+ \frac{g_S}{2} \sum_{a=0}^{8} \Bigl[ \left( \bar{q} \lambda^a q
\right)^2+ \left( \bar{q} (i \gamma_5)\lambda^a q \right)^2
 \Bigr] \nonumber \\
&+& g_D \Bigl[ \mbox{det}\bigl[ \bar{q} (1+\gamma_5) q \bigr]
  +  \mbox{det}\bigl[ \bar{q} (1-\gamma_5) q \bigr]\Bigr] \, .
\end{eqnarray}
The column vector  $q = (u,d,s)$ represents the quark field with three flavors, $N_f=3$,
and three colors, $N_c=3$. $\lambda^a$ are the Gell--Mann matrices, a = $0,1,\ldots , 8$,
${\lambda^0=\sqrt{\frac{2}{3}} \, {\bf I}}$.

The Lagrangian (\ref{lagr}) is invariant under chiral SU$_L(3)\otimes$SU$_R(3)$
transformations if we put $m_i=0$, where $m_i$ are the current quark masses
($\hat{m}=\mbox{diag}(m_u,m_d,m_s)$).
The last term in (\ref{lagr})  breaks the U$_A(1)$ symmetry. This term is a
reflection of the axial anomaly in QCD.

The model Lagrangian (\ref{lagr})  can be put in a form suitable for the
bosonization procedure after an adequate treatment of the last term, allowing
to obtain a four quark interaction from the six quark interaction.
Then the following effective quark Lagrangian is obtained:
\begin{eqnarray}
{\cal L}_{eff} &=& \bar q\,(\,i\, {\gamma}^{\mu}\,\partial_\mu\,-\,\hat m)\, q \,\,
+S_{ab}[\,(\,\bar q\,\lambda^a\, q\,)(\bar q\,\lambda^b\, q\,)]
+\,P_{ab}[(\,\bar q \,i\,\gamma_5\,\lambda^a\, q\,)\,(\,\bar q
\,i\,\gamma_5\,\lambda^b\, q\,)\,],
\label{lagr_eff}
\end{eqnarray}
where the projectors $S_{ab}\,, P_{ab}$ are given by:
\begin{eqnarray}
S_{ab} &=& g_S \delta_{ab} + g_D D_{abc}\left\langle \bar{q} \lambda^c q\right\rangle, \label{sab}\\
P_{ab} &=& g_S \delta_{ab} - g_D D_{abc}\left\langle \bar{q} \lambda^c q\right\rangle. \label{pab}
\end{eqnarray}
The constants $D_{abc}$ coincide with the SU(3) structure constants
$d_{abc}\,\,$ for $a,b,c =(1,2,\ldots ,8)$ and
$D_{0ab}=-\frac{1}{\sqrt{6}}\delta_{ab}$, $D_{000}=\sqrt{\frac{2}{3}}$.
The hadronization procedure can be done by the integration over the quark fields
in the functional integral with (\ref{lagr_eff}). The natural degrees of freedom
of low-energy QCD in the mesonic sector are achieved which gives the following
effective action:
\begin{align}
W_{eff}[\varphi,\sigma]  &  =-\frac{1}{2}\left(  \sigma^{a}S_{ab}^{-1}%
\sigma^{b}\right)  -\frac{1}{2}\left(  \varphi^{a}P_{ab}^{-1}\varphi
^{b}\right) \nonumber\\
&  -i\mbox{Tr}\,\mbox{ln}\Bigl[i\gamma^{\mu}\partial_{\mu}-\hat{m}%
+\sigma_{a}\lambda^{a}+(i\gamma_{5})(\varphi_{a}\lambda^{a})\Bigr]\,.
\label{action}
\end{align}
The notation $\mbox{Tr}$ stands for the trace operation over discrete indices
($N_{f}$ and $N_{c}$) and integration over momentum. The fields $\sigma^{a}$
and $\varphi^{a}$ are scalar and pseudoscalar meson nonets, respectively.

The first variation of the action (\ref{action}) leads to the gap equations,
\begin{eqnarray}\label{gap}
M_i = m_i - 2g_{_S} \big <\bar{q_i}q_i \big > -2g_{_D}\big <\bar{q_j}q_j\big > \big <\bar{q_k}q_k \big >\,,
\end{eqnarray}
with $i,j,k =u,d,s$ cyclic. $M_i$ are the constituent quark masses and the quark
condensates are given by: $\big <\bar{q}_i  q_i \big > = -i \mbox{Tr}[ S_i(p)]$, $S_i(p)$
being the quark Green function.

The baryonic thermodynamic potential of the grand canonical ensemble, $\Omega(T, V, \mu_i)$,
is also obtained directly from the effective action (\ref{action}). So we take the
temperature $T$, the volume $V$ and the chemical potential of the $i$-quark ($\mu_i$) as
the full independent state variables.

The relevant equations of state for the entropy $S$, the pressure $p$, and the particle
number $N_i$, as well as the internal energy $E$, follow from well known expressions like
the Gibbs-Duhem relation
\begin{equation}\label{tpot}
    \Omega (T, V, \mu_i )= E- TS - \sum_{i=u,d,s} \mu _{i} N_{i}\,.
\end{equation}
The following expressions are obtained:
\begin{eqnarray}\label{energy}
    E &=&- \frac{ N_c}{\pi^2} V\sum_{i=u,d,s}\left\{
    \int p^2 dp \, \frac{p^2 + m_{i} M_{i}}{E_{i}}\, (1\,-\,n_{i}-\bar n_{i})\right\}  \nonumber \\
    &&- g_{S} \sum_{i=u,d,s}\, (\big < \bar{q}_{i}q_{i}\big > )^{2}
    - 2 g_{D} \big < \bar{u}u\big > \big < \bar{d}d\big > \big < \bar{s}s\big > \,,
\end{eqnarray}
\begin{eqnarray}\label{entropy}
    S =-\frac{ N_c}{\pi^2} V \sum_{i=u,d,s}
    \int p^2 dp\,\,
    \biggl\{ \bigl[ n_{i} \ln n_{i}+(1-n_{i})\ln (1-n_{i})%
    \bigr] +\bigl[ n_{i}\rightarrow 1 - \bar n_{i} \bigr] \biggr\} \, ,
\end{eqnarray}
\begin{equation}\label{np}
    N_i = \frac{ N_c}{\pi^2} V \int p^2 dp\,\,
    \left( n_{i}-\bar n_{i} \right).
\end{equation}
V is the volume of the system and the quark density is determined by the relation $\rho_i =
N_i / V$. In the previous equations $n_i$ and $\bar n_i$  are the quark and antiquark
occupation numbers
\begin{equation}
n_{i}= \frac{1}{1+ e^{\beta(E_{i} - \mu_{i})}},\hskip1cm \bar n_{i} = \frac{1}{1+
e^{\beta(E_{i} + \mu_{i})}}.
\end{equation}

We define $\mu_B= \frac{1}{3} (\mu_u+\mu_d+\mu_s)$ and the baryonic matter density as
$\rho_B= \frac{1}{3} (\rho_u+\rho_d+\rho_s)$. As usual, the pressure and the energy density are
defined such that their values are zero in the vacuum state \cite{Buballa:2004PR}:
\begin{equation} \label{p}
    p (\mu_i, T) = - \frac{1}{V}\left[ \Omega(\mu_i, T) - \Omega(0, 0) \right] ,
\end{equation}
\begin{equation}\label{e}
    \epsilon(\mu_i, T) = \frac{1}{V}\left[E(\mu_i, T)-E(0,
    0)\right].
\end{equation}

The baryon number susceptibility is the response of the baryon number density $\rho_B(T,
\mu_i)$ to an infinitesimal variation of the quark chemical potential $\mu_i$
\cite{McLerran:1987PRD}:
\begin{equation} \label{chi}
    \chi_B = \frac{1}{3}\sum_{i=u,d,s}\left(\frac{\partial
    \rho_i}{\partial\mu_i}\right)_{T}.
\end{equation}

Another relevant observable, in the context of possible signatures for chiral symmetry
restoration in the hadron-quark transition and in transition from  hadronic matter to
the QGP \cite{McLerran:1987PRD,Asakawa:2000PRL,Blaizot:2001PLB}, is the specific heat
which is defined by \cite{PLBcosta}
\begin {equation}\label{c}
    C = \frac{T}{V}\left ( \frac{\partial S}{\partial T}\right)_{N_i}
    = \frac{T}{V}\left[\left ( \frac{\partial S}{\partial T} \right)_{\mu_i}
    - \frac{[(\partial N_i/\partial T)_{\mu_i}]^2}{(\partial N_i/\partial \mu_i)_T}\right],
\end {equation}
where we have transformed the derivative $(\partial S/\partial T)_{N_i}$ using the
formula of the Jacobian. In fact,  we work in the grand canonical  ensemble where
$(T,V,\mu_i)$ are the set of natural independent variables (still holding $N_i$ and $V$
fixed).

By expanding the effective action (\ref{action}) over meson fields, we get an effective
meson action from which we can obtain the meson propagators. In the present work we are
only concerned with $\pi^0$ and $\sigma$ mesons. Starting with the pseudoscalar mesons we
have the effective meson action:
\begin{equation}
W_{eff}^{(2)}[\varphi]=-\frac{1}{2}\varphi^{a}\left[  P_{ab}^{-1}-\Pi_{ab}%
^{P}(P)\right]  \varphi^{b}=-\frac{1}{2}\varphi^{a}(D_{ab}^{P}(P))^{-1}%
\varphi^{b}, \label{act2}%
\end{equation}
where $\Pi_{ab}^{P}(P)$ is the polarization operator,
\begin{equation}
\Pi_{ab}^{P}(P)=iN_{c}\int\frac{d^{4}p}{(2\pi)^{4}}\mbox{tr}_{D}\left[
S_{i}(p)(\lambda^{a})_{ij}(i\gamma_{5})S_{j}(p+P)(\lambda^{b})_{ji}%
(i\gamma_{5})\right],  \label{actp}
\end{equation}
with $\mbox{tr}_{D}$ is the trace over Dirac matrices. The expression in square brackets
in (\ref{act2}) is the inverse non-normalized meson propagator  $(D_{ab}^{P}(P))^{-1}$.

The inverse meson propagator for $\pi^0$ is given by
\begin{equation}
D^{-1}_{\pi^0} (P)= 1-P_{\pi^0} J_{uu}^P (P),
\end{equation}
with
\begin{equation}
P_{\pi^0}=g_{S}+g_{D}\left\langle\bar{q}_{s}q_{s}\right\rangle
\end{equation}
and where the polarization operator of the $\pi^0$ meson takes the form
\begin{equation}
J_{uu}^{P}(P_{0})=4\left[2I_{1}^{u}-P_{0}^{2}\,\,I_{2}^{uu}(P_{0})\right].
\label{ppij}
\end{equation}
The integrals $I_{1}^{i}$ and $I_{2}^{ij}(P_{0})$ are given in Appendix A.

The mass of the $\pi^0$ meson can be determined by the condition
$D_{\pi^0}^{-1}(M_{\pi^0},\mathbf{0})=0$ and the quark--meson coupling constant is
evaluated as
\begin{eqnarray}\label{mesq}
    g_{\pi^0\overline{q}q}^{-2} = -\frac{1}{2 M_{\pi^0}} \frac{\partial}{\partial P_0}
    \left[J_{uu}^{P}(P_0) \right]_{ \vert_{ P_0=M_{\pi^0}}}.
\end{eqnarray}

The procedure to describe scalar mesons is analogous.
We present below the most relevant steps.
Keeping now the scalar mesons only in (\ref{action}), we have the effective
meson action
\begin{equation}
W_{eff}^{(2)}[\sigma]=-\frac{1}{2}\sigma^{a}\left[  S_{ab}^{-1}-\Pi_{ab}%
^{S}(P)\right]  \sigma^{b}=-\frac{1}{2}\sigma^{a}({D}_{ab}^{S}(P))^{-1}%
\sigma^{b}, \label{accao2}%
\end{equation}
with $\Pi_{ab}^{S}(P)$ being the polarization operator, which in the momentum
space has the form of (\ref {actp}) with ($i\gamma_5$) substituted by the
identity matrix.

To consider the $\sigma$ meson we take into account the matrix structure of the
propagator in (\ref{accao2}). For the isospin symmetry considered in the present work
$\left\langle\bar{q}_{u}\,q_{u}\right\rangle=\left\langle\bar{q}_{d}\,q_{d}\right\rangle$,
and the matrices ${S}_{ab}$ and ${\Pi}_{ab}^{S}$ are reduced to
\begin{equation}
{S}_{ab}\rightarrow\left(
\begin{array}
[c]{cc}%
S_{33} & 0\\
0 & \bar{S}_{ab}%
\end{array}
\right)  \,\,\,\,\,\,\mbox{and}\,\,\,\,\,\,{\Pi}_{ab}^{S}\rightarrow\left(
\begin{array}
[c]{cc}%
\Pi_{33}^{S} & 0\\
0 & \bar{\Pi}_{ab}^{S}%
\end{array}
\right)  ,
\end{equation}
where the matrix elements are given in Appendix A.

The mass of the $\sigma$ meson can be determined by the condition
$D_{\sigma}^{-1}(M_{\sigma},\mathbf{0})=0$, where
\begin{equation}
D_{\sigma}^{-1}=\left(  \mathcal{A}+\mathcal{C}\right)  -\sqrt
{(\mathcal{C}-\mathcal{A})^{2}+4\mathcal{B}^{2}}%
\end{equation}
with $\mathcal{A}=S_{88}-\Delta\Pi^S_{00}(P),\,
\mathcal{C}=S_{00}-\Delta\Pi^S_{88}(P),\,\mathcal{B}=-(S_{08}+\Delta\Pi^S_{08}(P))$
and $\Delta=S_{00}S_{88}-(S_{08})^{2}$.

Finally, the model is fixed by the coupling constants $g_S$ and $ g_D$, the cutoff in
three-momentum space $\Lambda$, which is used to regularize the momentum space integrals
and the current quark masses $m_i$. For  numerical calculations in physical conditions we
use the parameter set \cite{Rehberg:1995PRC,Costa:2003PRC,Costa:2005PRD70,Costa:2005PRD71}:
$m_u = m_d =        5.5$ MeV,
$m_s =            140.7$ MeV,
$g_S \Lambda^2 =   3.67$,
$g_D \Lambda^5 = -12.36$ and
$\Lambda =        602.3$ MeV,
that has been determined by fixing the values
$M_{\pi} = 135.0$ MeV,
$M_K   = 497.7$ MeV,
$f_\pi =  92.4$ MeV, and
$M_{\eta'}= 960.8$ MeV.
For the quark condensates we obtain:
$\left\langle \bar{q}_{u}\,q_u\right\rangle = \left\langle\bar{q}_{d}\,q_d\right\rangle = - (241.9 \mbox{ MeV})^3$ and
$\left\langle\bar{q}_{s}\,q_s\right\rangle = - (257.7 \mbox{ MeV})^3$, and
for the constituent quark masses
$M_u= M_d= 367.7$ MeV and $M_s= 549.5$ MeV.


\section{Equations of state and  phase transition}\label{phase}

We will start the  discussion of the phase diagram of the NJL model (\ref{lagr}) by
analyzing the behavior of the pressure/energy per particle as a function of  the baryonic
density, paying special attention to the Gibbs criteria.
Our model of strong interacting matter can simulate either a region in the interior of a
neutron star or a hot and dense fireball created in a heavy-ion collision. In the present
work we focus our attention in the last type of systems, so we impose the condition
$\mu_e=0$, since electrons and positrons are not involved in the strong interaction. So,
we naturally get the chemical equilibrium condition $\mu_u=\mu_d=\mu_s=\mu_B$ that will
be used.  This choice allows for equal constituent quark masses $M_u=M_d$ and
approximates the physical conditions at RHIC.
In this respect, we remind that in a relativistic heavy-ion collision of duration
of $\sim 10^{-22}\, s$, thermal equilibration is possible only for processes mediated
by the strong interaction rather than the full electroweak equilibrium.

\begin{figure}
\begin{center}
  \begin{tabular}{cc}
    \hspace*{-0.5cm}\epsfig{file=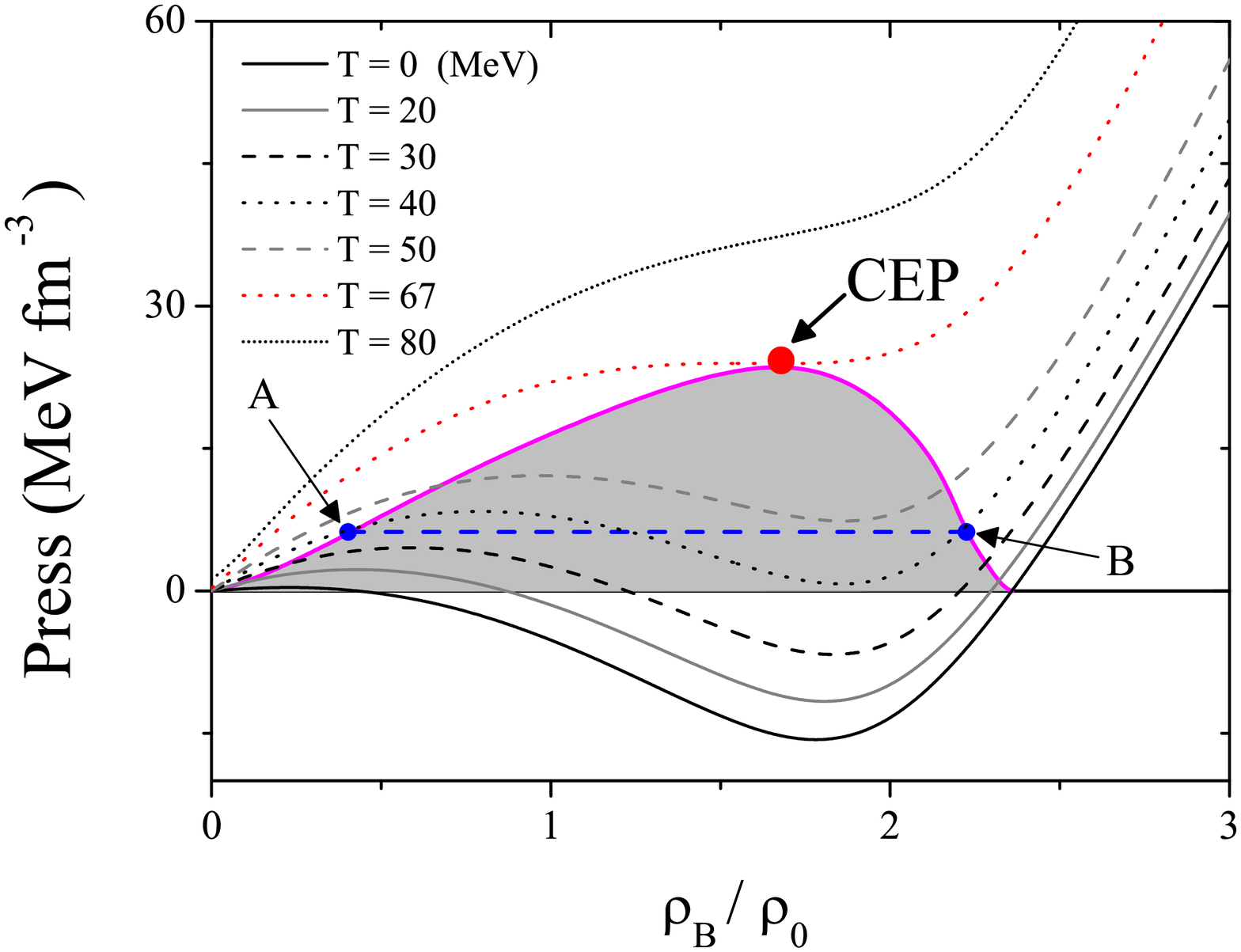,width=8.5cm,height=8cm} &
    \hspace*{-0.75cm}\epsfig{file=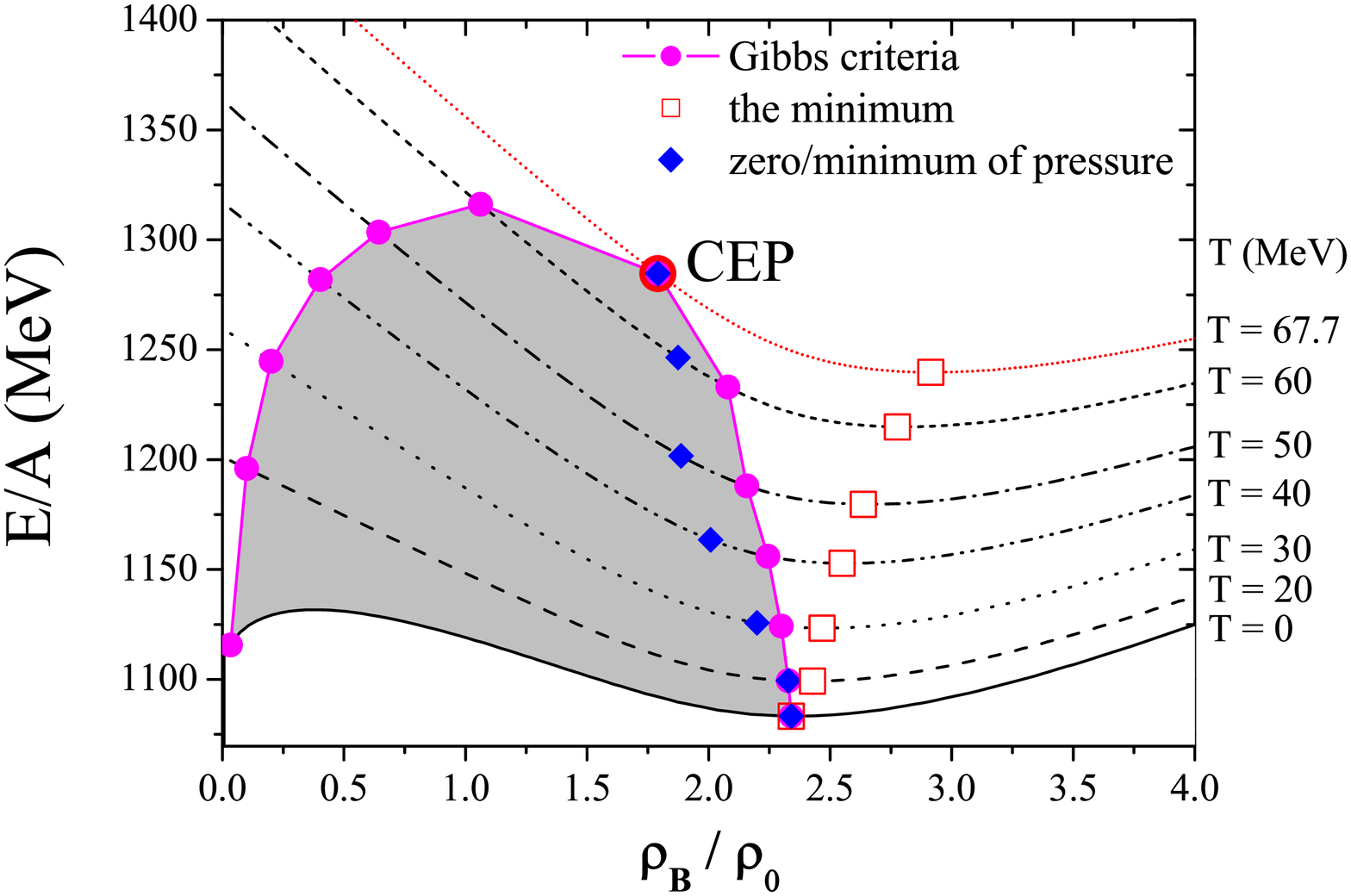,width=9cm,height=8cm} \\
   \end{tabular}
\end{center}
\vspace{-1.0cm} \caption{ Pressure (left) and energy  per particle
(right) as a function of the density at different temperatures. The
points $A$ and $B$ (left panel) illustrate the Gibbs criteria.
Only in the  $T=0$ line the zero-pressure point is located at
the  minimum of the energy per particle.}
\label{Fig:1}
\end{figure}

Let us  discuss  our results for the pressure/energy per baryon at zero temperature that
are plotted in Fig. \ref{Fig:1} as a function of $\rho_B/\rho_0$ (solid lines),
where  $\rho_0 = 0.17$ fm$^{-3}$ is the normal nuclear matter density. The pressure has
three zeros, respectively, at $\rho_B=0, 0.43 \rho_0, 2.36\rho_0$, that correspond to
the extreme of the energy per particle. For $\rho_B < 0.2 \rho_0$ the pressure and
compressibility are positive, so the system can exist in a uniform gas phase, but it will
not survive indefinitely, since the zero density state is energetically favored; for $
0.2 \rho_0<\rho_B < 0.43 \rho_0$ the system is unstable since the compressibility is
negative, in fact $\rho_B=0.43 \rho_0$ corresponds to a maximum of the energy per
particle; for $ 0.43 \rho_0<\rho_B < 2.36 \rho_0$, the pressure is negative, and the
third zero of the pressure, $\rho_B=2.36 \rho_0$, corresponds to an absolute minimum of
the energy (see Fig. \ref{Fig:1} (right panel)).
The appearance of an absolute minimum of the energy means the possibility for finite
droplets to be in mechanical equilibrium with the vacuum at zero -pressure ($P=0$).
Above $\rho_B=2.36 \rho_0$, which we define as $\rho_B^{cr}$, we have again a uniform gas
phase. So, for densities $0<\rho_B<\rho_B^{cr}$ the equilibrium configuration is a mixed
phase. This is because the Gibbs criterion of equal $P$ and $\mu_B$ is satisfied and,
therefore,  the phase transition  is a first order one: the thermodynamic potential has
two degenerate minima at which two phases have equal pressure and chemical potential and
can coexist. Such a situation is possible in regions where the gap equations have several
solutions for the quark masses.

Summarizing the results at $T=0$,  the behavior described allows the following
interpretation:  the uniform nonzero density  phase will break up into stable droplets
with zero pressure and density $\rho_B^{cr} = 2.36 \rho_0$ in which chiral symmetry is
partially restored, surrounded by a nontrivial vacuum with $\rho_B=P=0$ (see also
\cite{Buballa:NPA1998,Mish2000,Buballa:2004PR,Costa:2003PRC,Rajagopal:1999NPA}).
In fact, for our choice of the parameters the critical point at $T=0$ satisfies to the
condition $\mu_i<M_i^{vac}$ \cite{Buballa:2004PR,Scavenius}, where $M_i^{vac}$ is
the mass of the $i$-quark in the vacuum. This can be seen  by comparing
$\mu_B^{cr}=361$ MeV (see the T-axis of Fig. 2, left panel) with the quark masses
$M_u^{vac}\,=\,M_d^{vac}\,=\,367.7$ MeV and $M_s^{vac}\,=\,549.5$ MeV.

\begin{figure}
\begin{center}
  \begin{tabular}{cc}
    \hspace*{-0.5cm}\epsfig{file=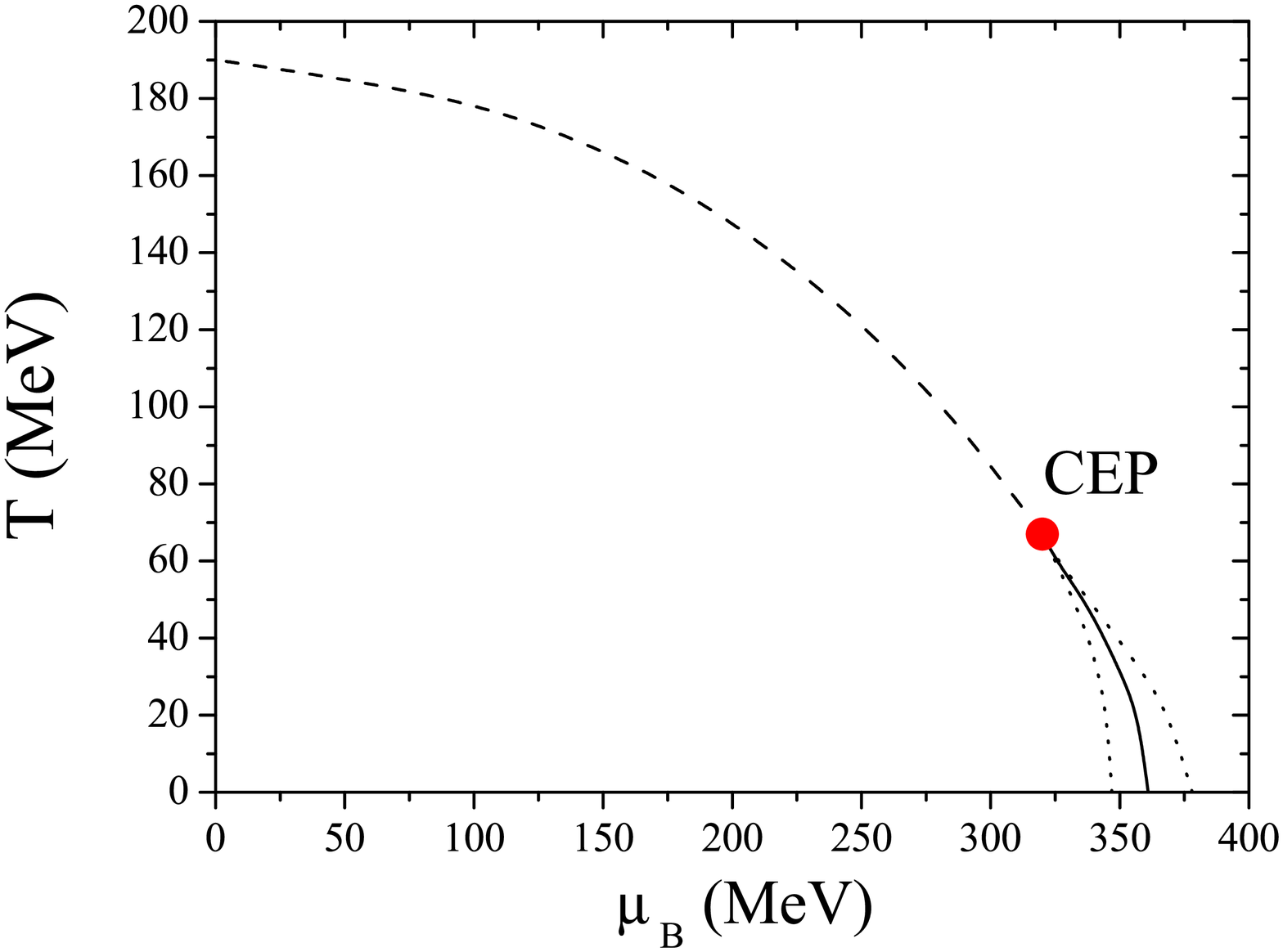,width=8.5cm,height=8cm} &
    \hspace*{-0.5cm}\epsfig{file=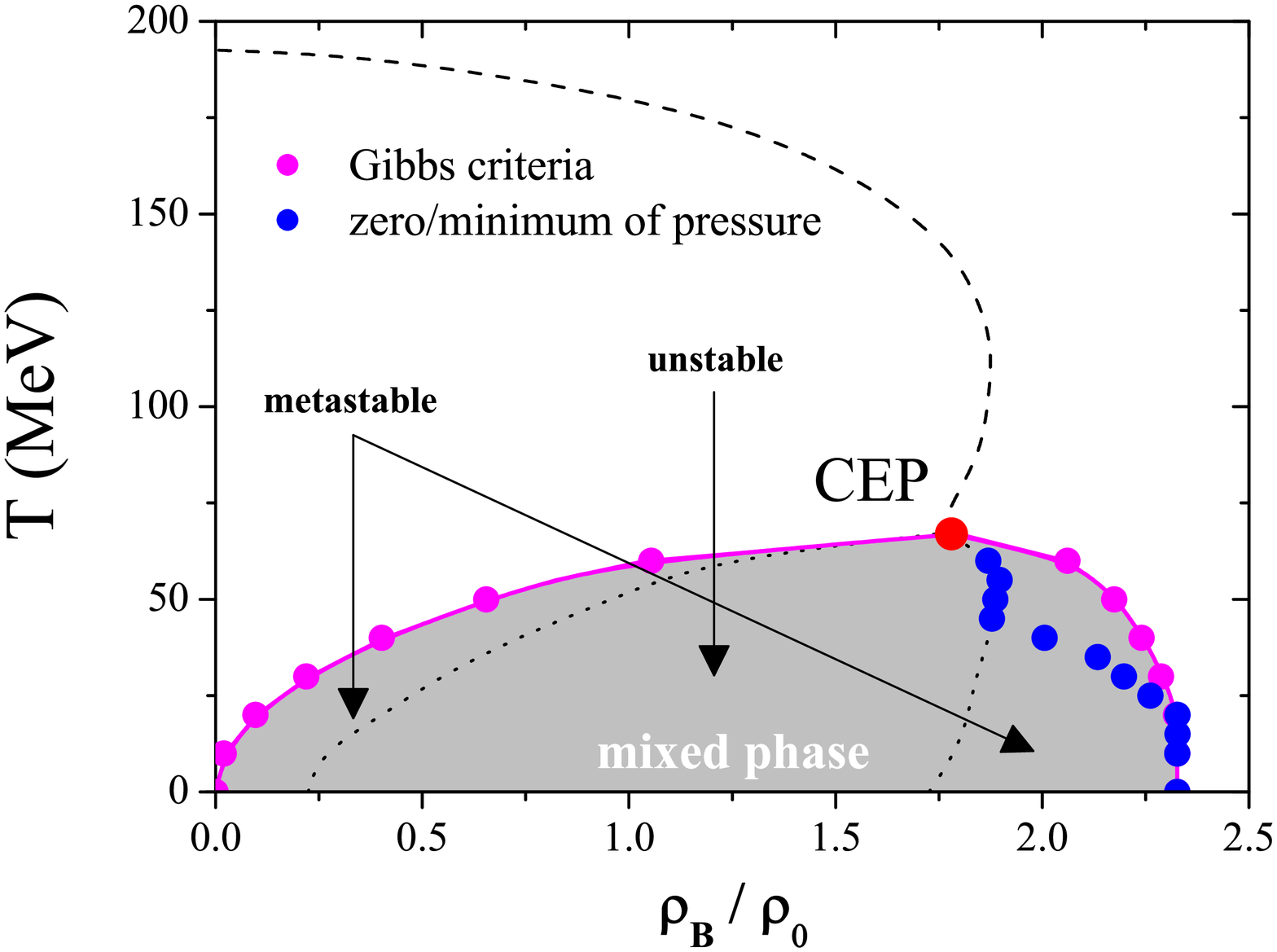,width=8.5cm,height=8cm} \\
   \end{tabular}
\end{center}
\vspace{-1.0cm} \caption{Phase diagram in the SU(3) NJL model. The left
(right) part corresponds to the $T-\mu_B$ ($T-\rho_B$) plane. Solid
(dashed) line shows the location of the first order (crossover)
transition. The dotted lines shows the location of the spinodal
boundaries of the two phase transitions (shown by shading in the
right plot).}
\label{Fig:2}
\end{figure}

As can be seen from Fig. \ref{Fig:1}, as the temperature increases, the first order
transition  persists up to the CEP. At the CEP the chiral transition becomes of second
order. Along the line of a first order phase transition the thermodynamic potential has
two degenerate minima. These minima are separated by a finite potential barrier making
the potential nonconvex. The height of the barrier is largest at zero temperature and
finite quark chemical potential and decreases towards higher temperature. At the CEP the
barrier disappears and the potential flattens. This pattern is characteristic of a first
order phase transition: the two minima correspond, respectively, to the phases of broken
and restored symmetry. The borders of the coexistence area are marked by the dotted lines
in  Fig. \ref{Fig:2}. The domain between the two dotted lines has metastable states which
are characterized by large fluctuations. They are also solutions  of the gap  equations
but their thermodynamic potential is higher than for the stable solutions. The left
dotted curves  represent the beginning of the metastable solutions of restored symmetry
in the phase of broken symmetry, while the right dotted curves represent the end of the
metastable solutions of broken symmetry in the restored symmetric phase. We also
represent in Fig. \ref{Fig:2} (right panel) the region where the solutions of the gap
equations are unstable.

The location of the CEP is found to be at $T^{CEP} = 67.7$ MeV and
$\rho_B^{CEP}=1.68\rho_0$ ($\mu_B^{CEP} = 318.4$ MeV). For
temperatures above the CEP the thermodynamic potential has only one
minimum and the transition is washed out: a smooth crossover takes
place.

Finally, we will focus again on the energy per baryon. In Fig. \ref{Fig:1}
(right panel), we plot the density dependence of the energy per baryon at different
temperatures.
We observe that the two points, zero of the pressure and  minimum of the energy density,
are not the same  at finite temperature. In fact, as can be seen from Fig.
\ref{Fig:1} (left panel), states with zero pressure are  only possible up to the
maximal temperature $T_m \sim 38$ MeV.
We notice that  the zero-pressure states persist up to temperatures of 70 MeV in a
two-flavor NJL  model where equal chemical potentials of quarks and antiquarks is assumed
\cite{Mish2000}.
For $T<T_m$ the zero-pressure states are in the metastable density region and, as soon as
$T\neq 0$, they do not coincide with the minimum of the energy per particle.

The  arguments just presented allow to clarify the difference between confined quark
matter (in hadrons) and bounded quark matter (droplets of quarks). As would be
expected, the binding mechanism is weaker than the confining one (nonexistent in the NJL
model). As a matter of fact, in spite of the existence of a binding energy for the
droplets of quarks at $T=0 $, we verify that it is not possible to avoid the evaporation
of the bounded quarks for arbitrarily small temperatures.

More detailed information concerning the structure of the phase
diagram will be given in Sec. V.


\section{Thermodynamic quantities in the $T-\mu_B$ plane}

For a better understanding of the thermodynamics of the phase transitions, we analyze in
this section the behavior of the  thermodynamical quantities that are the most relevant
to discuss the physics across the first order phase transition. With these quantities, we
can also discuss the latent heat which is inherent to this phase transition.

\begin{figure*}[t]
\begin{center}
  \begin{tabular}{cc}
    \hspace*{-0.5cm}\epsfig{file=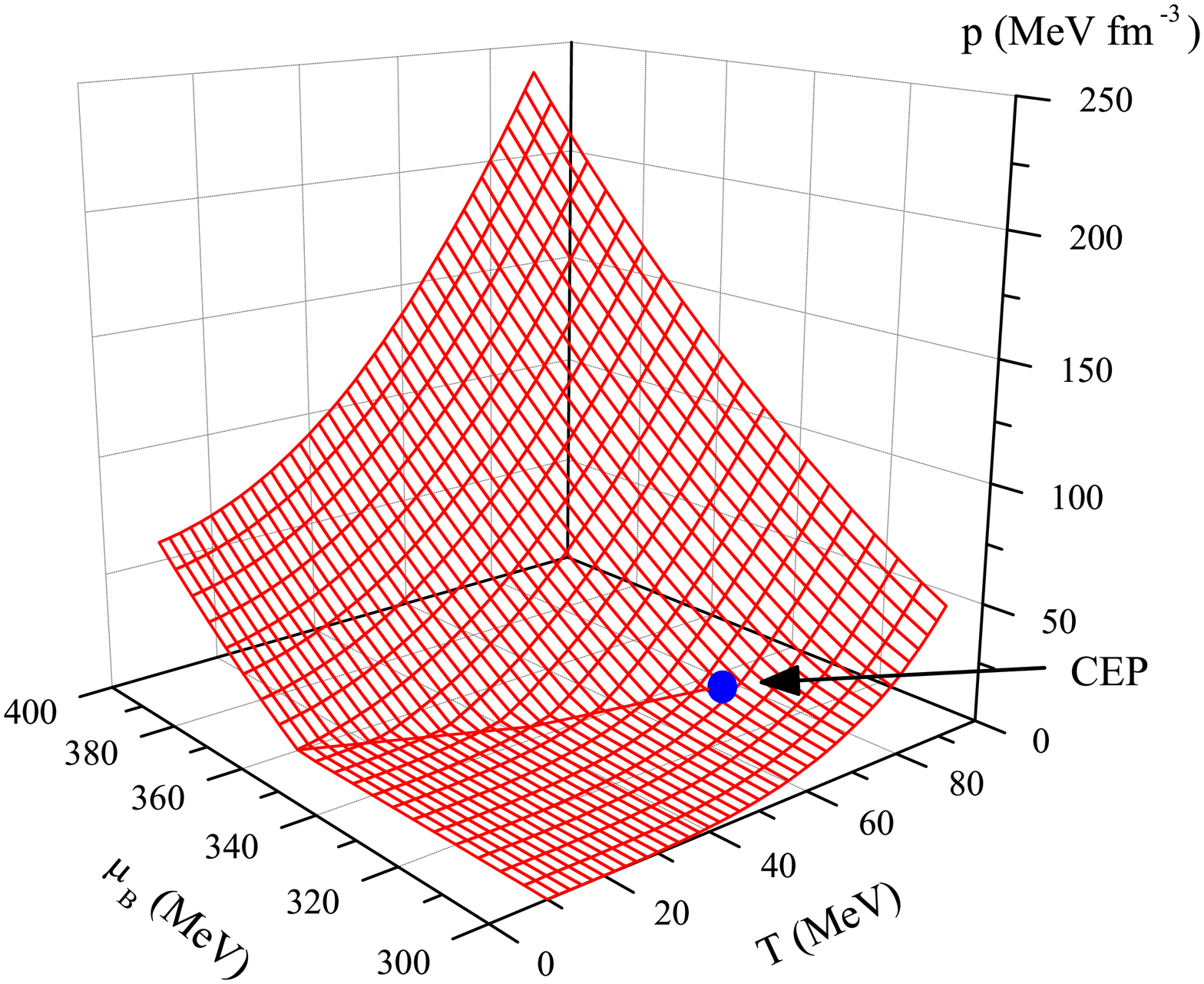,width=8.750cm,height=7.0cm} &
    \hspace*{-0.25cm}\epsfig{file=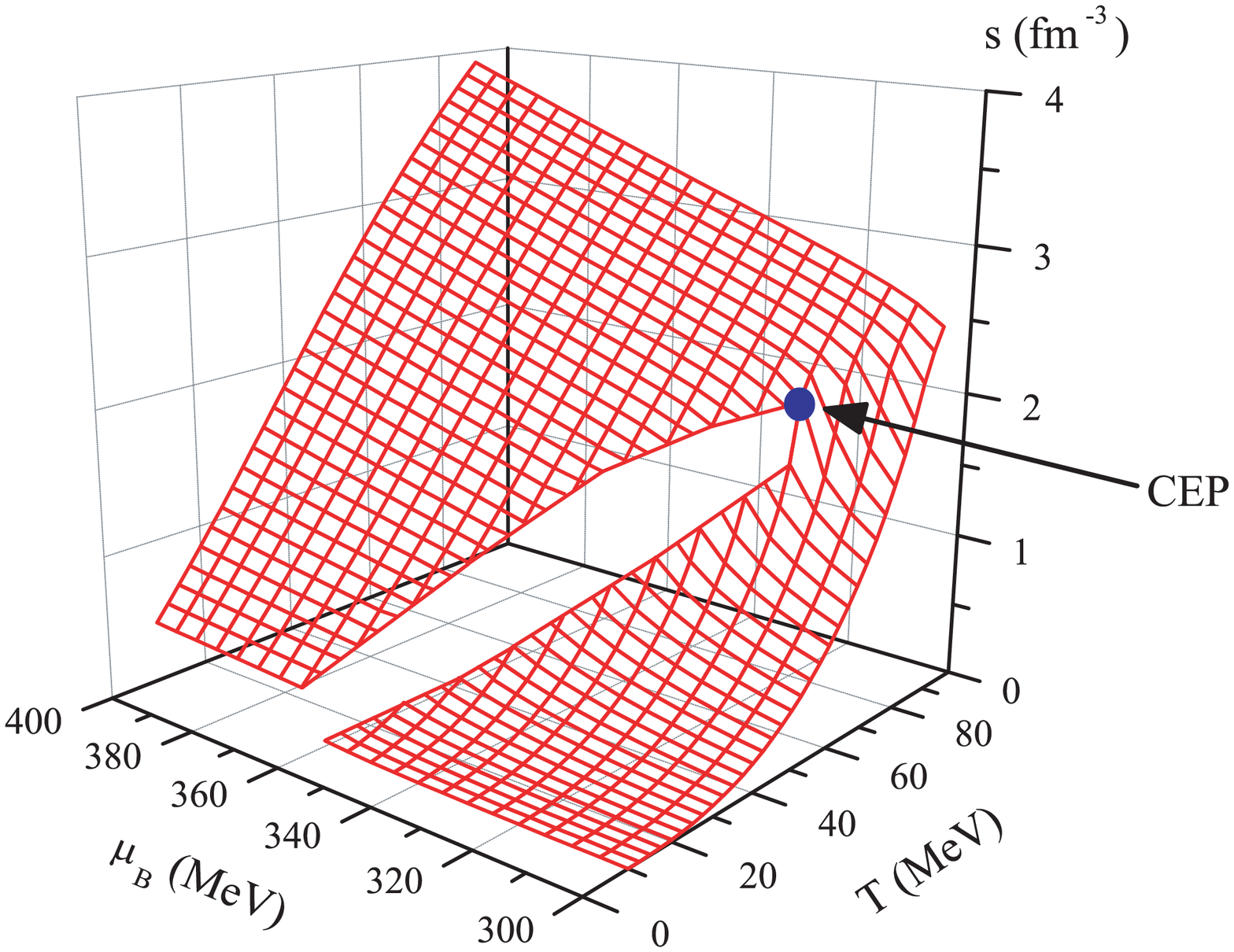,width=8.0cm,height=6.25cm} \\
   \end{tabular}
    \epsfig{file=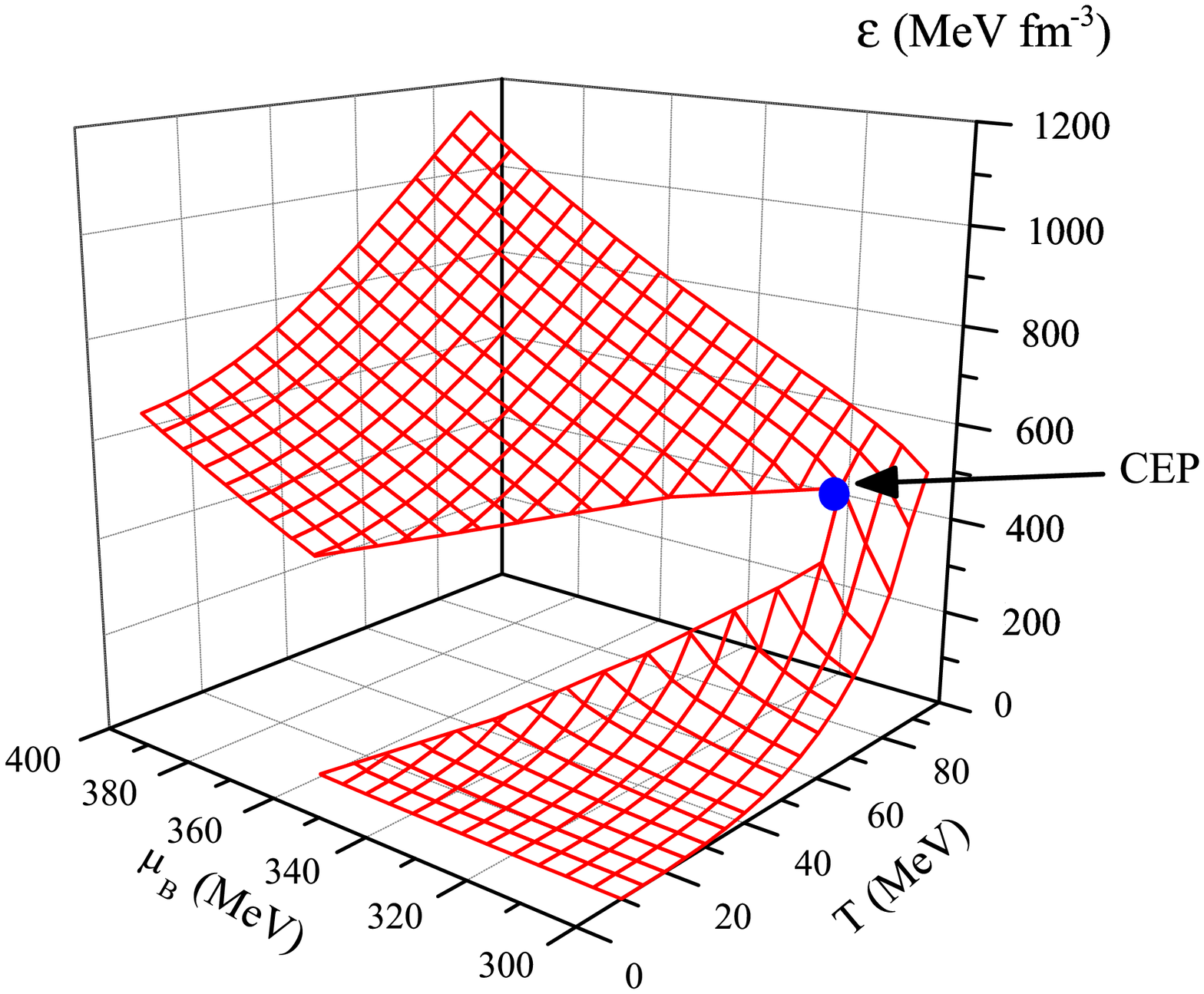,width=8.0cm,height=6.25cm}
\end{center}
\vspace{-0.5cm} \caption{Pressure (left side of upper part),   entropy  density (right
side of upper part) and energy density (down part) as functions of the temperature and
the baryonic chemical potential.}
\label{Fig:3}
\end{figure*}

The pressure is plotted in the left- hand side of Fig. \ref{Fig:3} (upper part), which
shows a continuous  behavior for all points of the phase diagram.
In a first order phase transition a discontinuity occurs in the  first derivatives of the
pressure (or the thermodynamic potential) with respect to $\mu_B$ and $T$, {\em i.e.},
the baryon number density and the entropy density, respectively.
In fact, as can be seen  in the right side of  Fig. \ref{Fig:3}, the entropy density
is discontinuous  in the first order phase transition region
($T<T^{CEP},\,\mu_B>\mu_B^{CEP}$). A similar behavior is found for the energy density,
whose curves show that the  first order phase transition, strong  at $T=0$, turns into  a
less abrupt one  as the temperature increases (see Fig. \ref{Fig:3}, lower part).
In the crossover transition ($T>T^{CEP},\,\mu_B<\mu_B^{CEP}$) the thermodynamic
quantities change rapidly within a narrow range of values of $T$ and $\mu_B$, but the
pressure and all its derivatives remain continuous, as shown in Fig. \ref{Fig:3}.

The discontinuities of the entropy and energy densities disappear at the CEP, which
location can not be determined  by universality arguments. The same is not true
concerning  local singular behavior of thermodynamic quantities around the CEP that will
be discussed in the next section through the critical exponents.

Let us now analyze what information concerning the latent heat we can get from our
results. As already referred in Sec. III, along the line of first order phase transition
the thermodynamic potential has two degenerate minima that are separated by a finite
barrier. This barrier is largest at zero temperature and finite chemical potential and
decreases towards higher temperature. At the CEP the barrier disappears, which  means
that there is no latent heat at this point.

As a grand canonical approach is applied to our model of strong interacting matter, the
independent quantities $T$ and $\mu_B$ represent the state variables which can be
externally controlled.
So, the conjugate of the intensive variables  $T$ and $\mu_B$  in the Legendre
transformation {---} the entropy density $s$ and the baryonic density $\rho_B$
{---} provide a more natural description.
By analyzing first the gap  in the curves of the entropy (Fig. \ref{Fig:3}, right side of
upper part), we see that the latent heat decreases for small temperatures, which is not
the expected behavior.
This analysis is, however, not  sufficient; both the baryonic density and the entropy
density contributions should be examined for more reliable information about the latent heat.
We remember that  the gap of the baryonic density across the first order phase transition
is largest at zero temperature and finite chemical potential and  vanishes at the CEP
(see Fig. \ref{Fig:2}, right panel).
The discontinuities in the energy density  include both  the entropy and the baryonic
density contributions and, as  can be seen in Fig. \ref{Fig:3}, they display the expected
behavior: the latent heat increases for decreasing temperatures.

Finally, to understand  the thermodynamics of matter created  in relativistic heavy-ion
collisions, it is convenient to calculate thermodynamic quantities along lines of
constant entropy per baryon number, the so-called isentropic lines. Most of these studies
have been done on lattice calculations for two-flavor QCD at finite $\mu_B$
\cite{Ejiri1200}, where nonphysical mass spectrum that corresponds to a too large of pion
mass $m_\pi \simeq 770$ MeV, has been used.  Such studies predict  that the effects of
the CEP change only slowly as the collision energy is changed as a consequence of the
attractor character of the CEP \cite{Stephanov:1998PRL}.

\begin{figure}[t]
\begin{center}
  \begin{tabular}{cc}
    \hspace*{-0.5cm}\epsfig{file=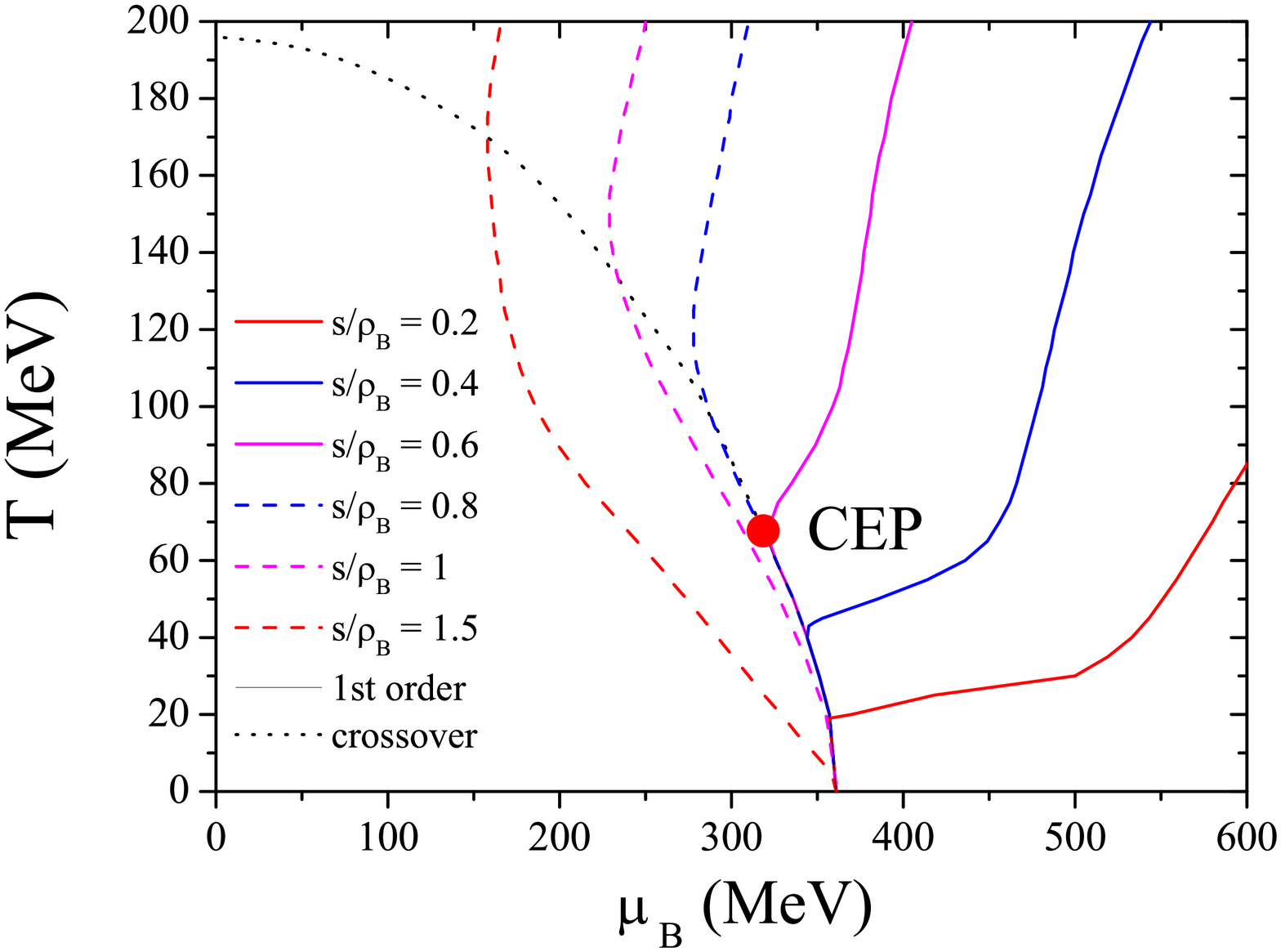,width=8.50cm,height=8cm} &
    \hspace*{-0.75cm}\epsfig{file=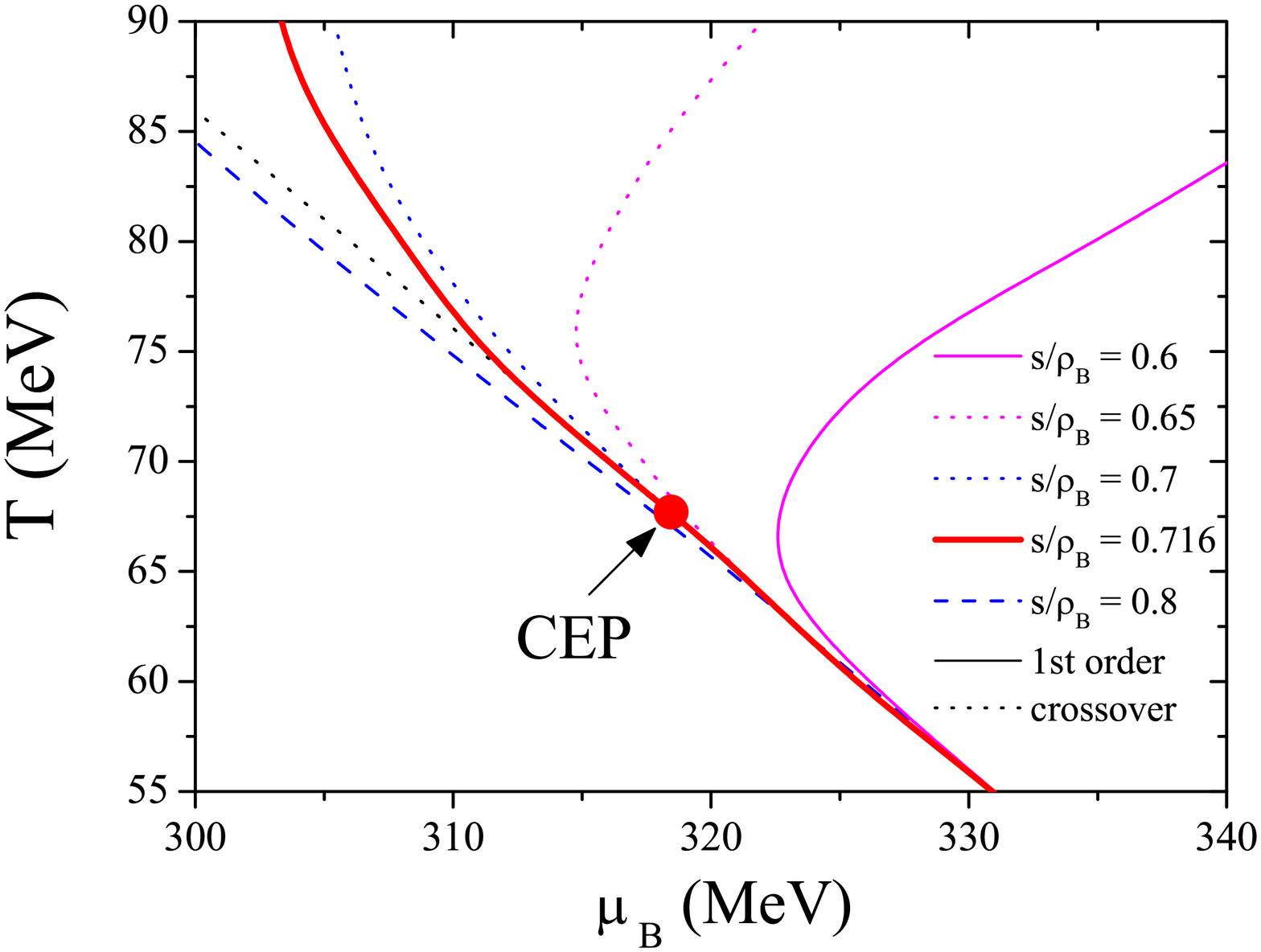,width=8.50cm,height=8cm} \\
   \end{tabular}
\end{center}
\vspace{-0.5cm}
\caption{Two perspectives of the entropy per baryon number in the
$T-\mu_B$ plane. The vicinity of the CEP is enlarged in the right panel.}
\label{Fig:4}
\end{figure}

Our model calculations for the isentropic lines in the $T-\mu_B$ plane are shown in Fig.
\ref{Fig:4}. The behavior we find is somewhat different from those claimed by other
authors \cite{Stephanov:1999PRD,Ejiri1200,Nonaka}, where a phenomena of focusing of
trajectories towards the CEP is observed.

The isentropic trajectories in the phase diagram (Fig. \ref{Fig:4}) indicate that the
slope of the trajectories goes to large values for large $T$. This behavior is related to
the effects of the strange quark mass in our model. In fact,  at high  temperatures the
relation $\mu_s>M_s$ is verified, allowing for a more pronounced decrease of $M_s$
\cite{Costa:2003PRC}.
Although  the entropy and the baryon number density, at high temperatures, are sensitive
to the regularization procedure used \cite{Zhuang:1994NPA,Costa:2007}, this effect is not
relevant for the present situation. The same is not true with respect to the effects of
the value of the cutoff itself  in the regime of low temperatures as will be shown below.

In a small range of $s/\rho_B$ around $0.7$ (see Fig. \ref{Fig:4}, right panel), we observe a
tendency of convergence of these isentropic
lines towards the CEP. These lines come from the region of
symmetry partially restored in the direction of the  crossover line.
For smaller values of $s/\rho_B$, the isentropic lines  turn about the CEP and then
attain the first order transition line. For larger values of $s/\rho_B$ the isentropic
trajectories approach the CEP by the region where the chiral symmetry is still broken,
and also attain the first order transition line after bending toward the critical point.
As already pointed out in \cite{Scavenius}, this is a natural result in these type of
quark models with no change in the number of degrees of freedom of the system in the two
phases. As the temperature decreases a first order phase transition occurs, the latent
heat increases and  the formation of the mixed phase is thermodynamically favored.

Finally, we remark that all isentropic  trajectories directly terminate at $T=0$ at the
first order transition line, without reheating in the mixed phase as verified in the
"zigzag" shape of  \cite{Subramanian,Stephanov:1999PRD,Ejiri1200,Nonaka}.
It is also  interesting to point out that, in the limit $T\rightarrow 0$, it is verified
that $s \rightarrow 0$ and $\rho_B \rightarrow 0$, as it should be.
This behavior  is in  contrast to \cite{Scavenius} (right panel Fig. 9) using the
NJL model in the SU(2) sector and is related  to our more convenient choice of the model
parameters, mainly a lower  value of the cutoff. This can be explained by the presence of
droplets at $T=0$ whose stability is quite sensitive to the choice of the model parameters.
In fact, as referred in  Sec. III, our choice of the parameters has  important effects:
we verify that, at $T=0$, the phase transition connects the vacuum state ($P=0,
\rho_B=0$) directly with the phase of chiral symmetry partially restored ($P=0,
\rho=\rho_B^{cr}$) and   the critical point of the phase transition in these conditions
satisfies to $\mu_i<M_i^{vac}$, where $M_i^{vac}$ is the mass of the $i$-quark
($i=u,d,s$) in the vacuum. This condition fulfills  the criterium of stability of the
quark droplets \cite{Buballa:2004PR,Costa:2003PRC}. In addition, it is also crucial to
the satisfaction of the third law of thermodynamics in the limit $T\rightarrow 0$. This
cutoff effect has an identical role in the formation of stable droplets on both SU(2) and
SU(3) NJL models.


\section{Phase diagrams and susceptibilities  in the vicinity of the
critical end point}

In this section we analyze with more detail the phase diagrams  in different conditions
in the  $T-\mu_B$ plane. Lattice QCD calculations have established the transition to a
phase where quarks and gluons are deconfined at temperatures larger than $\sim 150$ MeV
and zero baryon density. Depending on the number of quark flavors $N_f=2$ or $N_f=3$, and
on the masses of the quarks, different situations can occur and the transition from
hadronic matter to QGP may be of first order, second order, or a crossover transition.
To confront the model results with the universality arguments, we will discuss the class
of the critical points by changing the current quark masses in SU(2) and SU(3) versions
of the NJL model.


\subsection{Characteristics of the $T-\mu_B$ phase diagram}

\begin{figure}[t]
\begin{center}
  \begin{tabular}{cc}
    \hspace*{-0.5cm}\epsfig{file=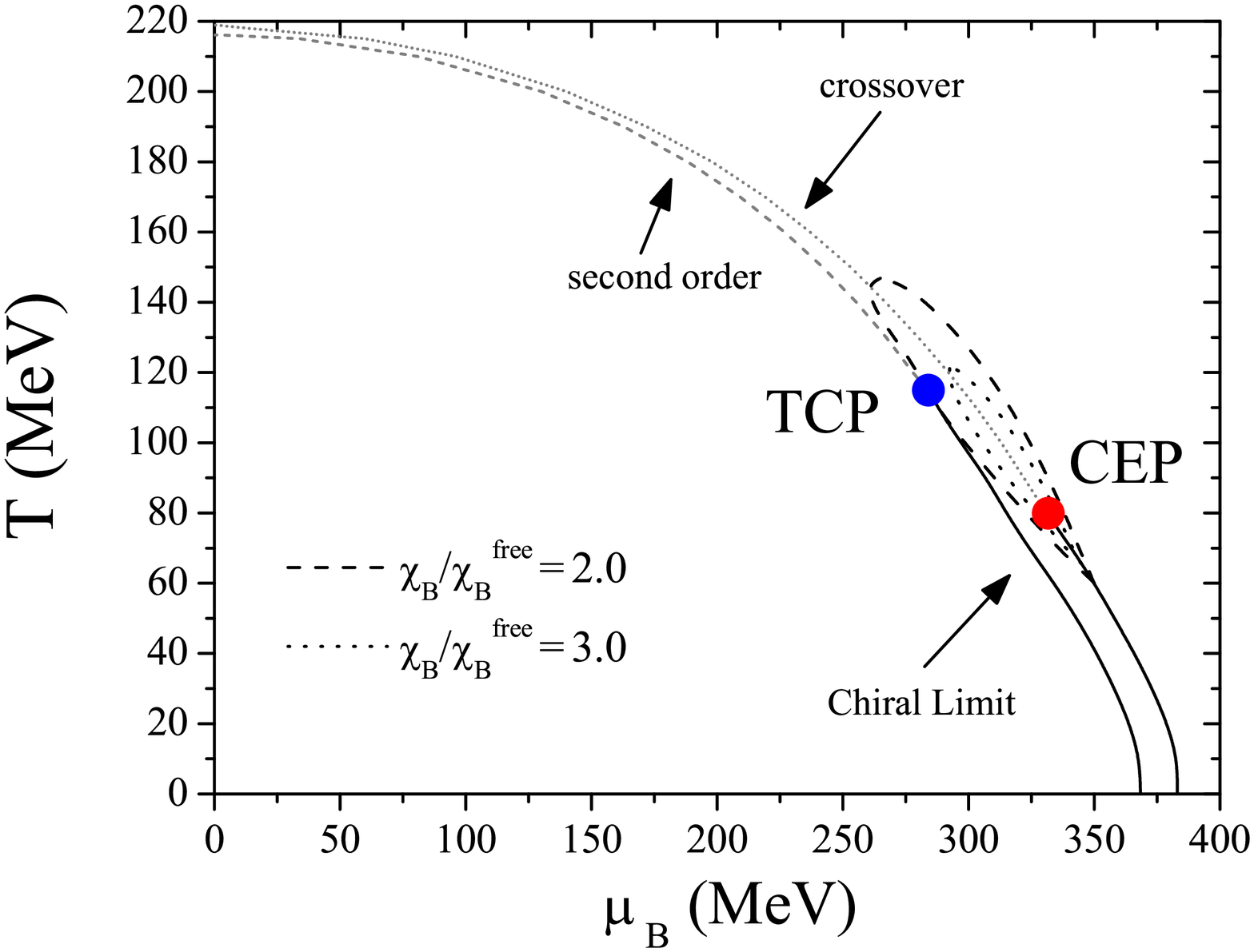,width=8.50cm,height=8cm} &
    \hspace*{-0.75cm}\epsfig{file=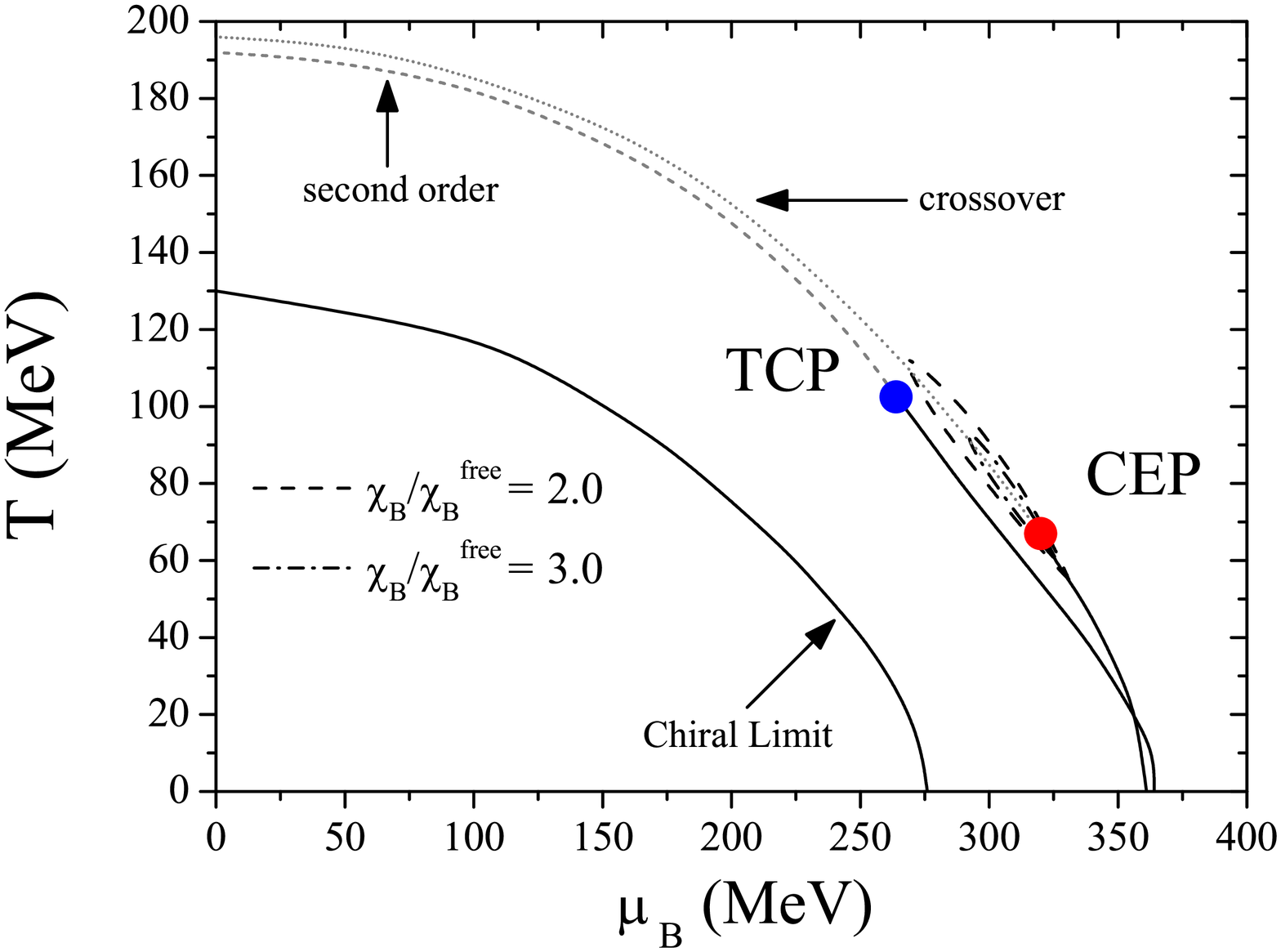,width=8.50cm,height=8cm} \\
   \end{tabular}
\end{center}
\vspace{-0.5cm} \caption{Phase diagram in the SU(2) (left) and SU(3) (right) NJL models.
The solid line represents the first order phase transition, the dashed line the
second order phase transition, and the dotted line the crossover transition.
The size of the critical region is also plotted for several values of
$\chi_B/\chi_B^{free}$. The TCP in the right panel is found for $m_u=m_d=0$ MeV and
$m_{s}=140.7$ MeV.}
\label{Fig:5}
\end{figure}

We start by analyzing  the differences between the three-flavor NJL model and its simpler
version in the SU(2) sector. The phase diagrams for both models  are  presented in Fig.
\ref{Fig:5} as a function of $\mu_B$ and $T$.

Concerning the SU(2) model, and  using physical values of the quark masses: $m_u = m_d =
5.5$ MeV, we find that the CEP is localized at $T^{CEP}=79.9$ MeV and $\mu_B^{CEP} =
331.72$ MeV ($\rho_B^{CEP}=2.26 \rho_0$).
We also verified that, in the chiral limit,  the transition is of second order at
$\mu_B=0$ and, as $\mu_B$ increases, the line of  second order phase transition will end
in a first order line at the TCP. The TCP is located at $\mu_B^{TCP}=286.1$ MeV and
$T^{TCP}=112.1$ MeV.

For the SU(3) NJL model, also in the chiral limit ($m_u=m_d=m_s=0$), we verify that the
phase diagram does not exhibit a TCP: chiral symmetry is restored via a first order
transition for all baryonic chemical potentials and temperatures (see right panel of Fig.
\ref{Fig:5}).
According to lattice analysis, this pattern of chiral symmetry restoration should  remain
even when the strange quark acquires a nonzero   current mass, provided it is  lower
than a critical value ($m_s < m_s^{crit}$), and $m_u=m_d=0$ is still kept. The value  for
$m_s^{crit}$ is  not settled yet,  those found   in lattice \cite{Laermann}  or in model
calculations \cite{Hsu:1998PLB,Barducci:2005PRD} being  lower than the physical strange
current quark mass  ($m_s\approx 150$ MeV). We found $m_s^{crit}=18.3$ MeV in our model
\cite{PLBcosta}, lower than lattice values \cite{Laermann} but consistent with what it is
expected in these type of models \cite{Barducci:2005PRD}.

When $m_s\geq m_{s}^{crit}$, at $\mu_B=0$, the
transition is of second order and, as $\mu_B$ increases, the line of the second order
phase transition will end in a first order line at the TCP.  The TCP for $m_{s}=140.7$
MeV is the closest to the CEP \cite{PLBcosta} and is located at $\mu_B^{TCP}=265.9$ MeV
and $T^{TCP}=100.5$ MeV. If we choose $m_u=m_d\neq0$, instead of second order transition
we have a smooth crossover whose critical line will end in the first order line at the
CEP.
Using physical values for the quark masses
\cite{Rehberg:1995PRC,Costa:2005PRD70}: $m_u = m_d = 5.5$ MeV, $m_s = 140.7$ MeV, this
point is localized at $T^{CEP}=67.7$ MeV and $\mu_B^{CEP} = 318.4$ MeV
($\rho_B^{CEP}=1.68\rho_0$).

We point out that both situations are in agreement with what is expected at $\mu_B=0$:
the chiral phase transition at the chiral limit is of second order for $N_f = 2$ and
first order for $N_f\geq3$ \cite{Pisarski:1984PRD}.

We also observe that the critical region is heavily stretched in the direction of the
crossover transition line, in both $N_f=2$ and $N_f=3$ cases,  as shown in Fig.
\ref{Fig:5}.
To estimate the critical region around the CEP we  calculate the dimensionless ratio
$\chi_B/\chi_B^{free}$, where $\chi_B^{free}$ is the chiral susceptibility of a free
massless quark gas.
The left (right) panel of Fig. \ref{Fig:5} shows a contour plot
for two fixed ratios $\chi_B/\chi_B^{free}=2.0;3.0$ in the phase diagram around the CEP.


\subsection{Behavior of $\chi_B$ and $C$  in the vicinity of the critical end point
and their critical exponents}

The phenomenological relevance of fluctuations in the finite temperature and chemical
potential around the CEP/TCP of QCD has been recognized by several authors.
If the critical region of the CEP is small, it is expected that most of the fluctuations
associated with the CEP will come from the mean field region around the
CEP \cite{Hatta:2003PRD}.
The size of the critical region around the CEP can be found by calculating the baryon
number susceptibility, the specific heat and their critical behaviors.

To a better understanding of the critical behavior of the system, we also  analyze in
some detail what happens in the SU(2) case, sector to which there is more information in
the literature \cite{Sasaki}.

As is well known, the  baryon number susceptibility, $\chi_B$, and the specific heat,
$C$, diverge at $T = T^{CEP}$ \cite{Hatta:2003PRD,Schaefer:2006,PLBcosta}. In order to
make this statement more precise, we will focus on the values of a set of indices, the
so-called critical exponents, which describe the behavior near the critical point of
various quantities of interest (in our case $\epsilon$ and $\alpha$ are the critical
exponents of $\chi_B$ and $C$, respectively). The motivation for this study arises from
fundamental phase transition considerations, and thus transcends any particular system.
These critical exponents will be determined by finding two directions, temperature
and magnetic-field-like, in the $T-\mu_B$ plane near the CEP, because, as pointed out in
\cite{Griffiths:1970PR}, the strength of the divergence is governed by the critical
exponents whose values depend on the path approaching the CEP.

\begin{table}[t]
\begin {center}
\begin{tabular}{cccccccc}
    \hline
    {Quantity} & {  critical exponents/path} && { SU(2) NJL} && { SU(3) NJL} && {Universality} \\
    \hline \hline
  & {$\epsilon\,/\,\,\rightarrow$\textcolor{red}{$\bullet$}} && { {$0.66 \pm 0.01$}}
  && {$0.67 \pm 0.01$} && {$2/3$} \\
  {$\chi_B$} & {{$\epsilon^\prime$\,/\,\,\textcolor{red}{$\bullet$}$\leftarrow$}} &&
  {$0.66 \pm 0.01$} && {$0.68 \pm 0.01$} && {$2/3$} \\
  &  $\gamma_B\,/\rightarrow$\textcolor{blue}{$\bullet$} &&
  $0.51 \pm 0.01$ && $0.49 \pm 0.02$ &&  {$1/2$}  \\
    \hline
    & {$\alpha\,/
        \begin{array}{c}
          \textcolor{red}{\bullet} \\
          \uparrow
        \end{array}$}
    && {$
        \begin{array}{c}
          \alpha=0.59\pm 0.01 \\
          \alpha_1=0.45\pm 0.01
        \end{array}$}
    && {$
        \begin{array}{c}
          0.61\pm 0.01 \\
          $---$
        \end{array}$}
    && {$
        \begin{array}{c}
        2/3 \\
        $---$
        \end{array}$} \\
  {$C$}
  & {$\alpha^\prime/
        \begin{array}{c}
          \downarrow \\
          \textcolor{red}{\bullet}
        \end{array}$}
    && {$0.69 \pm 0.01$} && {$0.67 \pm 0.01$} && {$2/3$} \\
    & {$\alpha\,/
        \begin{array}{c}
        \textcolor{blue}{\bullet} \\
        \uparrow
        \end{array}$}
    && {$0.40 \pm 0.01$} && {$0.45 \pm 0.02$} && {$1/2$}  \\
\hline
\end{tabular}

\begin{flushleft}
TABLE I: The arrow $\rightarrow\textcolor{red}{\bullet}$  $\left(
\begin{array}{c}
  \textcolor{blue}{\bullet} \\
  \uparrow
\end{array}
\right)$
indicates the path in the $\mu_B\,(T)-$ direction to the CEP (TCP)
for ${\mu_B<\mu_B^{CEP}}$ $({T<T^{TCP}}$).
\end{flushleft}

\end{center}
\label{tab:(I)}
\end{table}

Considering  the baryon number susceptibility, if the path chosen is asymptotically
parallel to the first order transition line at the CEP, the divergence of $\chi_B$ scales
with an exponent $\gamma_B$. In the mean field approximation it is expected that $\gamma_B=1$
for this path. For directions not parallel to the tangent line the divergence scales as
$\epsilon =2/3$. These values  are responsible for the elongation of the critical region,
$\chi_B$, being enhanced in the direction parallel to the first order transition line (see
Fig. \ref{Fig:5}).

To study the critical exponents for the baryon number susceptibility (Eq. \ref{chi}) we
will start with a path parallel to the $\mu_B$-axis in the $T-\mu_B$ plane, from lower
$\mu_B$ towards the critical $\mu_B^{CEP}$, at fixed temperature $T = T^{CEP}$.  Using a
linear logarithmic fit
\begin{equation}
    \ln \chi_B = -\epsilon \ln |\mu_B -\mu_B^{CEP} | + c_1 ,
\end{equation}
where the term $c_1$ is independent of $\mu_B$, we obtain $\epsilon = 0.67\pm 0.01$,
which is consistent with the mean field theory prediction $\epsilon = 2/3$.

We also study the baryon number susceptibility from higher $\mu_B$ towards the
critical $\mu_B^{CEP}$. The logarithmic fit used now is
$\ln \chi_B = -\epsilon' \ln |\mu_B -\mu_B^{CEP}| + c'_1$.
Our result shows that $\epsilon' = 0.68\pm 0.01$ which is very near the value
of $\epsilon$. This means that the size of the region we observe is approximately
the same independently of the direction we choose for the path parallel to the
$\mu_B$-axis.
These critical exponents,  calculated  in both SU(2) and SU(3) NJL models,
are presented in Table I.

For comparison purposes with the universality/mean field predictions, the calculated
critical exponents at the TCP  are also presented in Table I.
It is  found that
the critical exponent for $\chi_B$, $\gamma_B$ once we are in the TCP,  has the  value
$\gamma_B=0.49\pm0.02$, for the SU(3) NJL model and $\gamma_B=0.51\pm0.01$, for the SU(2)
NJL model. These results  are in agreement with the  mean field value ($\gamma_B=1/2$),
and show that the behavior of the baryon number susceptibility is similar in both SU(2)
and SU(3) versions of the model.

\begin{figure}[t]
\begin{center}
  \begin{tabular}{cc}
    \hspace*{-0.5cm}\epsfig{file=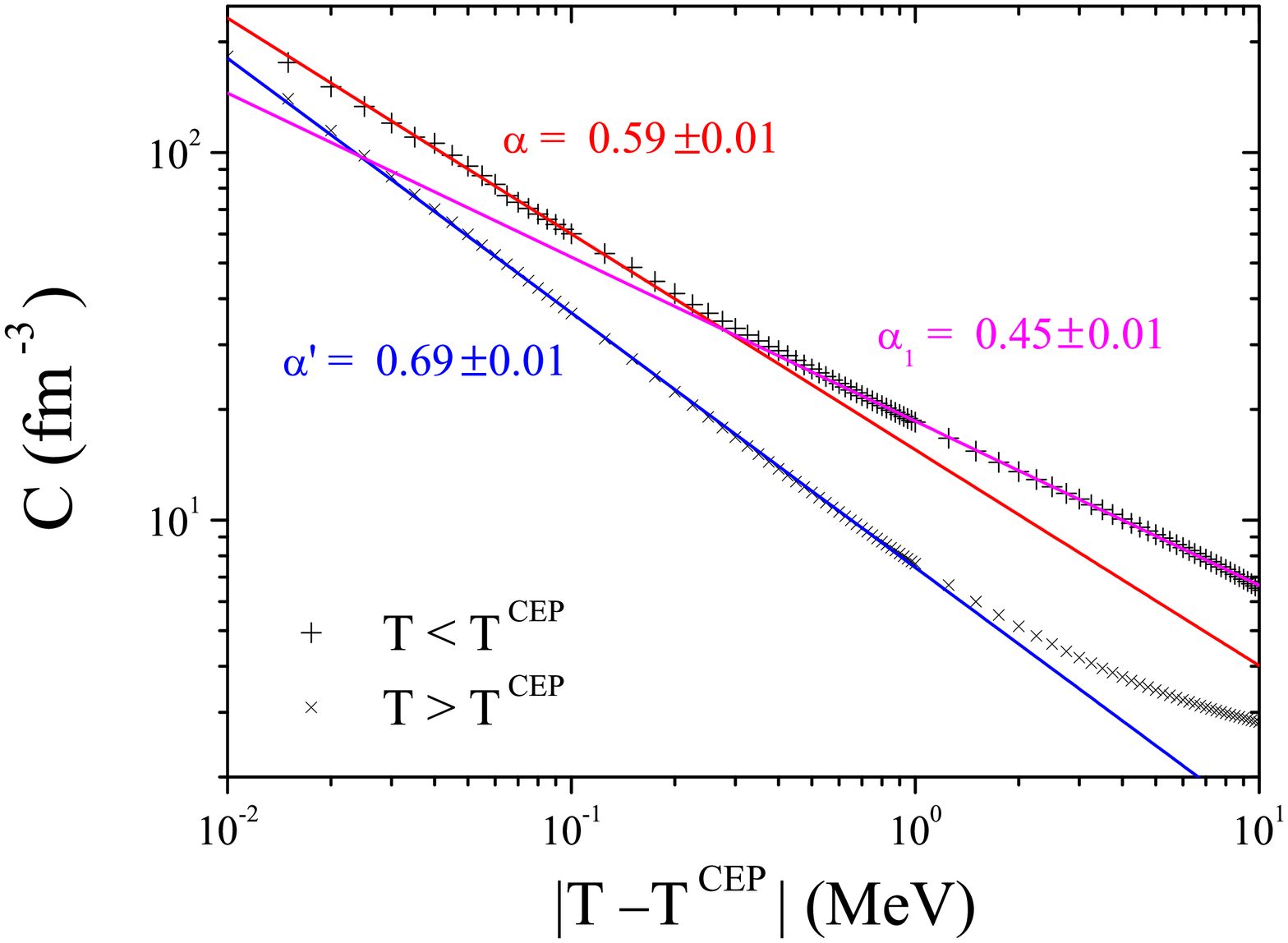,width=8.50cm,height=8cm} &
    \hspace*{-0.75cm}\epsfig{file=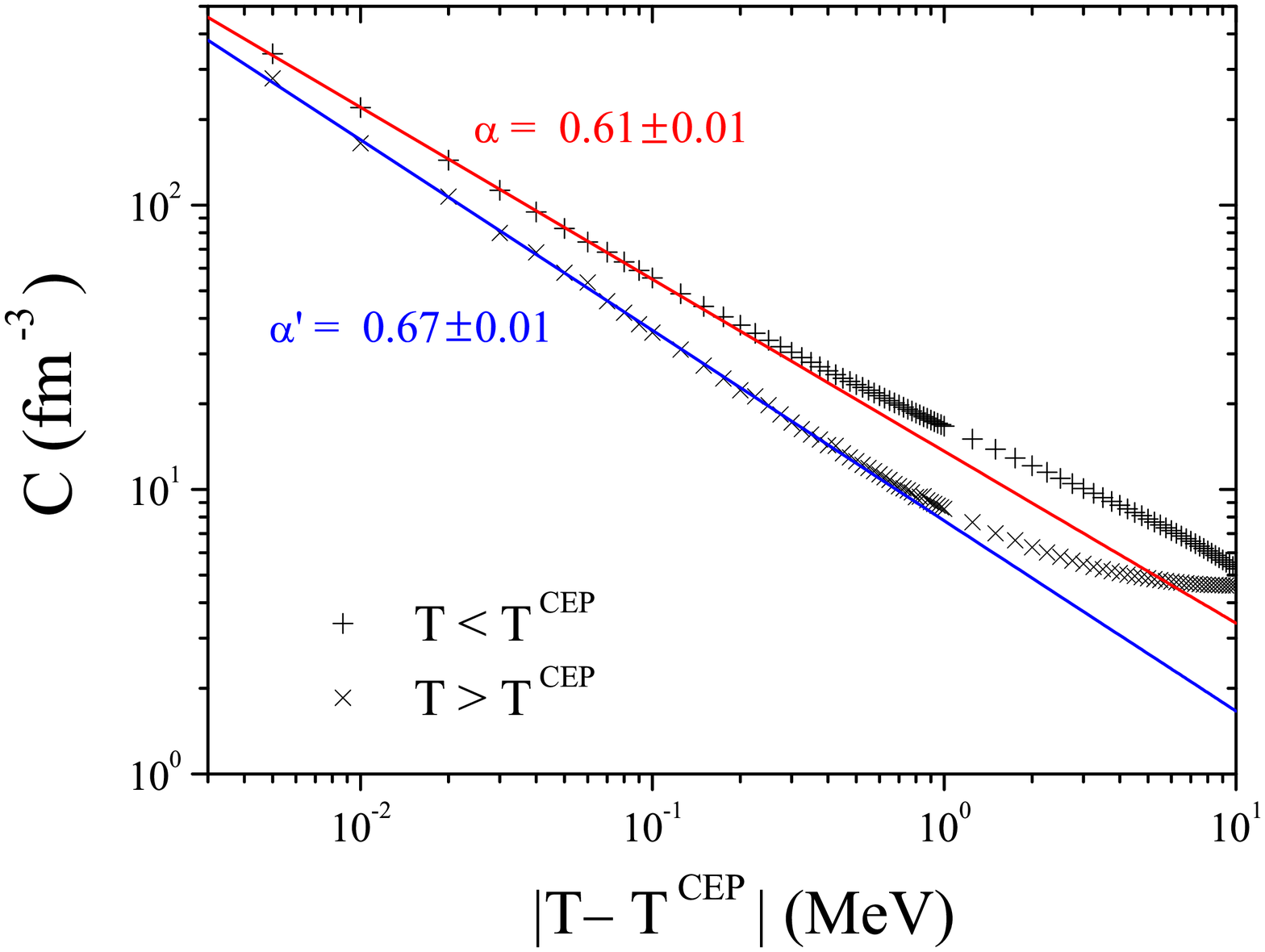,width=8.50cm,height=8cm} \\
    \hspace*{-0.5cm}\epsfig{file=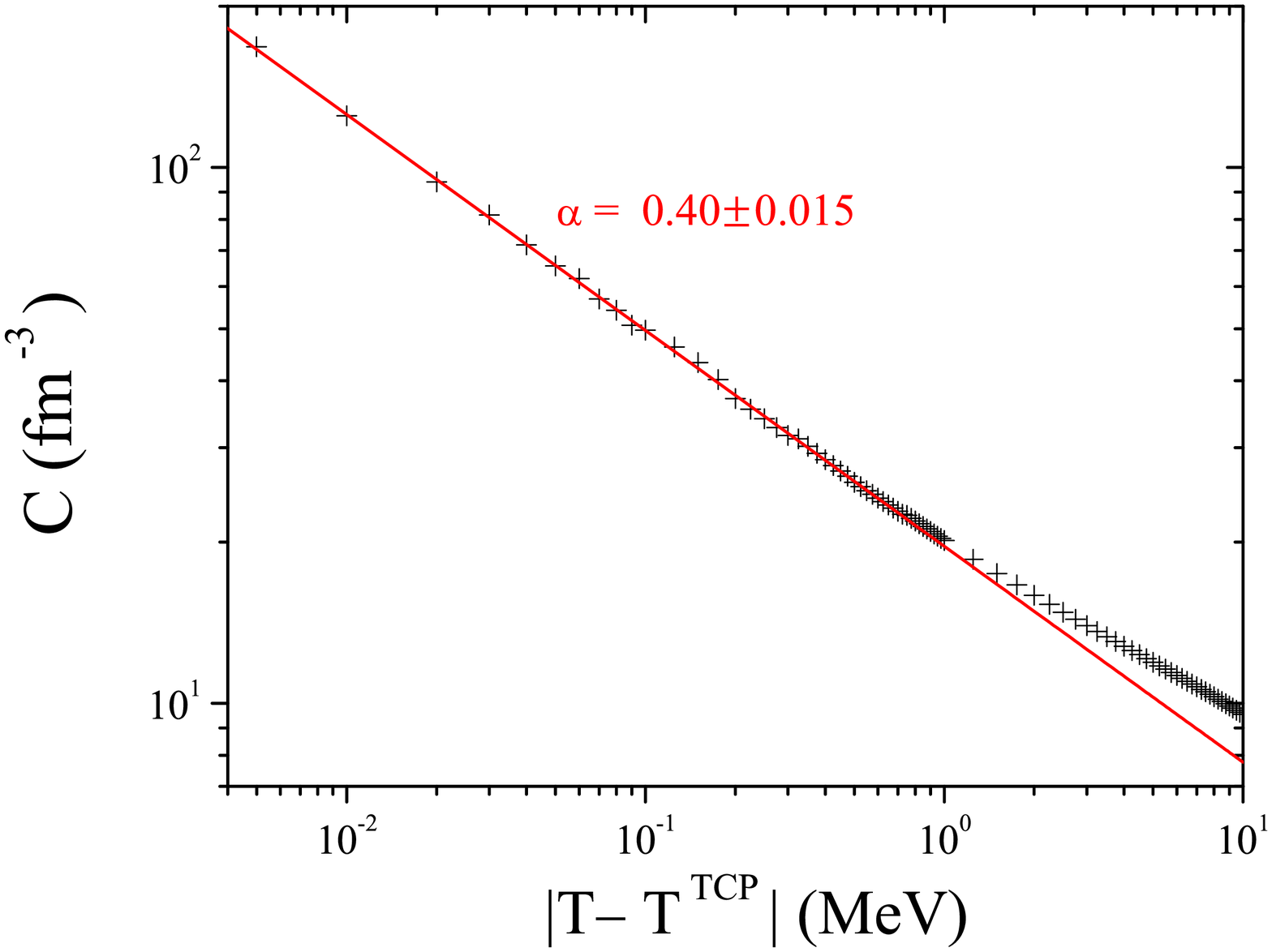,width=8.50cm,height=8cm} &
    \hspace*{-0.75cm}\epsfig{file=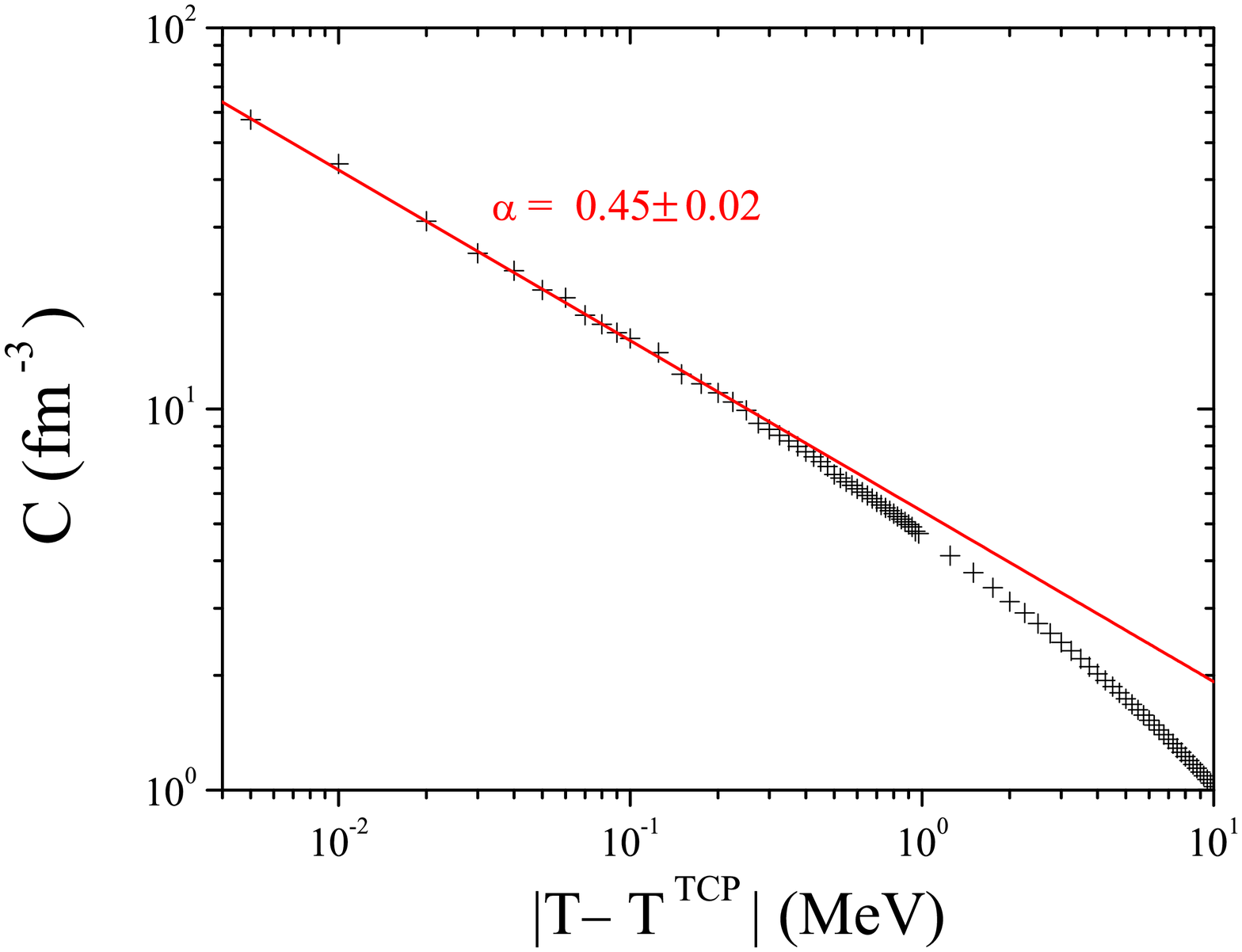,width=8.50cm,height=8cm} \\
   \end{tabular}
\end{center}
\vspace{-0.5cm} \caption{  Upper part: Specific heat as a function of $|T-T^{CEP}|$ at
the fixed chemical potential $\mu^{CEP}_B$ for SU(2) (left) and SU(3) (right) NJL models.
Lower part: Specific heat as a function of $|T-T^{TCP}|$ at the fixed chemical potential
$\mu^{TCP}_B$ for SU(2) (left) and SU(3) (right) NJL models. }
\label{Fig:6}
\end{figure}

Paying now attention to the specific heat (Eq. \ref{c}) around the CEP, we have used a
path parallel to the $T$-axis in the $T-\mu_B$ plane from lower/higher $T$ towards the
critical $T^{CEP}$ at fixed $\mu_B = \mu_B^{CEP}$. In Fig. \ref{Fig:6} (upper part) we
plot $C$ as a function of $T$ close to the CEP in a logarithmic scale for both SU(2) and
SU(3) calculations.
In this case we use the linear logarithmic fit $\ln C = -\alpha \ln |T -T^{CEP}| + c_2$,
where the term $c_2$ is independent of $T$.
Starting with the SU(2) case, we observe in the left panel that, for the region
$T<T^{CEP}$, we have a slope of data points that changes for values of $|T-T^{CEP}|$
around $0.3$ MeV. We have fitted the data for $|T-T^{CEP}|<0.3$ MeV and $|T-T^{CEP}|>0.3$
MeV separately and obtained, respectively, the critical exponent $\alpha=0.59\pm 0.01$
and $\alpha_1=0.45\pm 0.01$, which have a linear behavior for several orders of magnitude
(see also Table I). As pointed out in \cite{Hatta:2003PRD}, this change of the exponent
can be interpreted as a crossover of different universality classes, with the CEP being
affected by the TCP. It seems that in our model the effect of the hidden TCP on the CEP
is relevant for the specific heat contrarily to what happens to $\chi_B$.

We also observe that there is no clear evidence of the change of the slope of the fitting
of data points  in the three-flavor NJL model (see Fig. \ref{Fig:6}, right panel of
upper part, and Table I).
In fact, now we  only  obtain  a critical exponent
$\alpha=0.61\pm 0.01$ when the critical point is approached from below. When the critical
point is approached from above the trivial exponent  $\alpha^\prime =0.67\pm 0.01$ is
obtained.

To explore the possible effect  of the hidden TCP on the CEP, as suggested in Refs.
\cite{Hatta:2003PRD,Schaefer:2006}, we analyze the behavior of the specific heat around
the TCP.
As shown in Fig. \ref{Fig:6} (lower part) and Table I, we find nontrivial critical exponents
$\alpha=0.40\pm0.01$ and $\alpha=0.45\pm0.01$,  for SU(2) and SU(3) cases, respectively.
This result, in spite of being close, is not in agreement with the respective mean field value
($\alpha=1/2$).
However, they can justify the crossing effect observed.
We notice that the closest distance between the TCP and the CEP in the phase diagram
occurs in the T-direction ($(T^{TCP} - T^{CEP})<(\mu_B^{CEP} - \mu_B^{TCP})$).

The inconsistency with the mean field values only occurs for the exponent $\alpha$
as can be seen from Table I.
According to what was  suggested by universality arguments in \cite{Hatta:2003PRD}, it
was expected that $\chi_B$ and $C$ should be essentially the same near the TCP and the
CEP, which would imply $\alpha=\epsilon=2/3$ at the CEP.
Nevertheless we observe that the nontrivial values of $\alpha$ in the TCP and in the CEP
are consistent within the NJL model for both, SU(2) and SU(3) versions of the model, and they
reflect the effect of the TCP on the CEP. We also stress that the universality arguments
are so general that they give no quantitative results and, due to the lack of information
from the lattice simulations, they should be confronted with model calculations.
Our results seem particularly interesting because the NJL model shares with QCD some
features, such as the dynamics of chiral symmetry.
In particular, the physics underlying the critical singularities in the QCD diagram is
associated with this fundamental property of strong interaction.
So, the NJL model is an useful framework allowing for insights to the difficult task of
the analysis of the QCD phase diagram at finite temperature and chemical potential.

The eventual difference between the values of the $C$ and $\chi_B$
critical exponents can be interesting in heavy-ion collisions experiments.


\section{Partial and effective restoration of chiral symmetry}

As we have shown in previous sections, thermodynamics provides a well- established
procedure, as for instance the Gibbs criterion, to determine the critical points for the
phase transition  in the first order region. It follows that these critical points are
signalized by the discontinuity of several relevant observables (masses, quark
condensates) at some critical chemical potential, a situation that does not happen in the
crossover region, where these observables are continuous. At present, the criterion  most
commonly accepted, and that will be used here, to define the critical point in the
crossover region,  is to identify this point as  the inflection point of the quark masses
$\partial^{2}M/\partial T^{2}=0$ \cite{Buballa:2004PR} or, equivalently, of the quark
condensates \cite{Alles,Ruivo05}. This criterion is numerically equivalent  to the one
first proposed by  M. Asakawa and K. Yazaki that defines the point   where  the
constituent quark masses decrease to  half of their values in the vacuum
($M_{u}=M_{u}(0)/2$) \cite{Asakawa:1989NPA}, as the critical point. From this point on
the quark masses decrease quickly.

Both in the first order and in the crossover regions it is verified that  the quark
masses, especially for the  non strange quarks, decrease strongly at the critical point.
However, at this point different observables violating chiral symmetry are still far from
zero, like the quark condensates, the pion decay constant, and the difference between the
masses of the chiral partners.
One can say, therefore, that at the critical point  there occurs only a \textit{partial}
restoration of chiral symmetry.

In view of what was said above we use the following criteria: we define the point in the
$T- \mu_B$ plane for the phase transition associated with \textit{partial} restoration of
chiral symmetry as the inflexion (discontinuity) point for the quark masses, and define
the point for \textit{effective} restoration of chiral symmetry as that one where the
masses of chiral partners become degenerate.
This is also signaled by the merging of the $\pi^0$ and $\sigma$ spectral functions
\cite{Hubert}.

As we include the strange sector in this study, the consequences of the nonvanishing
anomaly term (mixing effects) on the strangeness content of mesons and mixing angles must
be analyzed. In fact, as the temperature (density) increases, the mixing angles get
close to their ideal values and the strangeness content of the mesons change 
\cite{Costa:2005PRD70,Costa:2005PRD71}: the masses of the mesons that become almost 
non-strange, $\sigma$ and $\eta$, converge, respectively, with those of the non strange 
mesons $\pi^0$ and $a_0$, while that of the $\eta^\prime$,  that becomes essentially 
strange, does not get close to  $f_0$ (see \cite{Costa:2005PRD71} (Fig. 2, Case I)); 
the convergence of the chiral partner $(\kappa, K)$, that has a $\bar u s$ structure, 
occurs at higher temperatures and is probably slowed by the small decrease of the 
constituent strange quark mass, $M_s$. 
For the purpose of discussing the {\em effective} restoration of chiral symmetry, 
we restrict our analysis to the chiral partners $(\pi^0, \sigma$) that behave in a 
qualitatively similar manner as the pair $(a_0, \eta)$.

The behavior of the masses of the  chiral partners ($\pi^{0},\sigma$) in the limiting
cases ($T\neq 0$, $\rho_B=0$) and ($T=0$, $\rho_B\neq 0$) are qualitatively similar and
well known from the literature: they both converge at a certain value of the temperature
(density). The main difference between the finite temperature and the finite density case
is that, in the first one, the degeneracy of the chiral partners occurs in a range of
temperatures where the mesons are no longer bound states: the $\pi^{0}$ dissociates in
$q\bar{q}$ pair at the Mott temperature $T_{M\,\pi^0}=212$ MeV
\cite{Rehberg:1995PRC,Costa:2003PRC}, and  the $\sigma$ at the Mott temperature
$T_{M\,\sigma}=160$ MeV; for the finite density case, the mesons are always bound states.

\begin{figure}[t]
\begin{center}
  \begin{tabular}{cc}
    \hspace*{-0.5cm}\epsfig{file=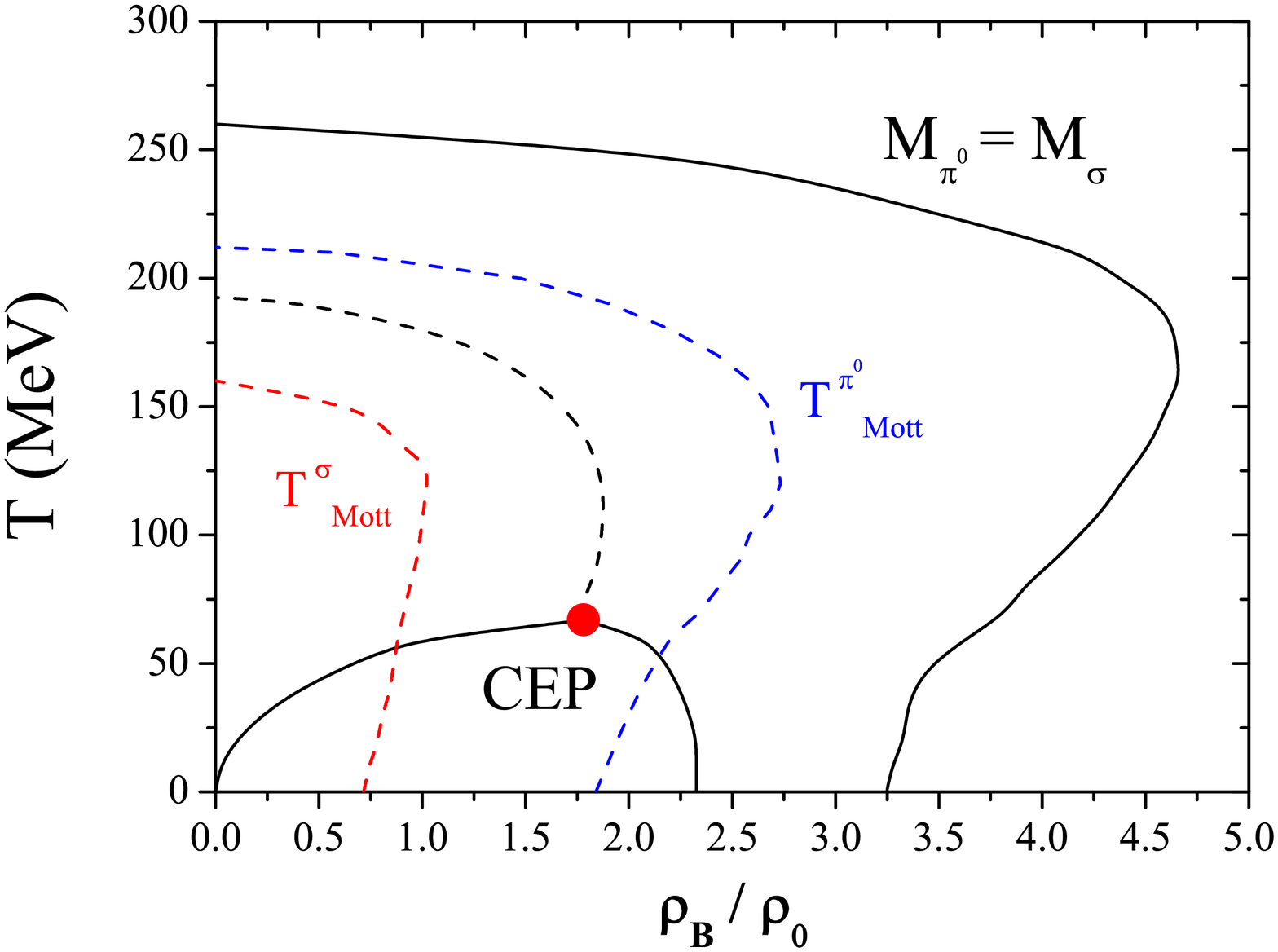,width=8.5cm,height=8cm} &
    \hspace*{-0.5cm}\epsfig{file=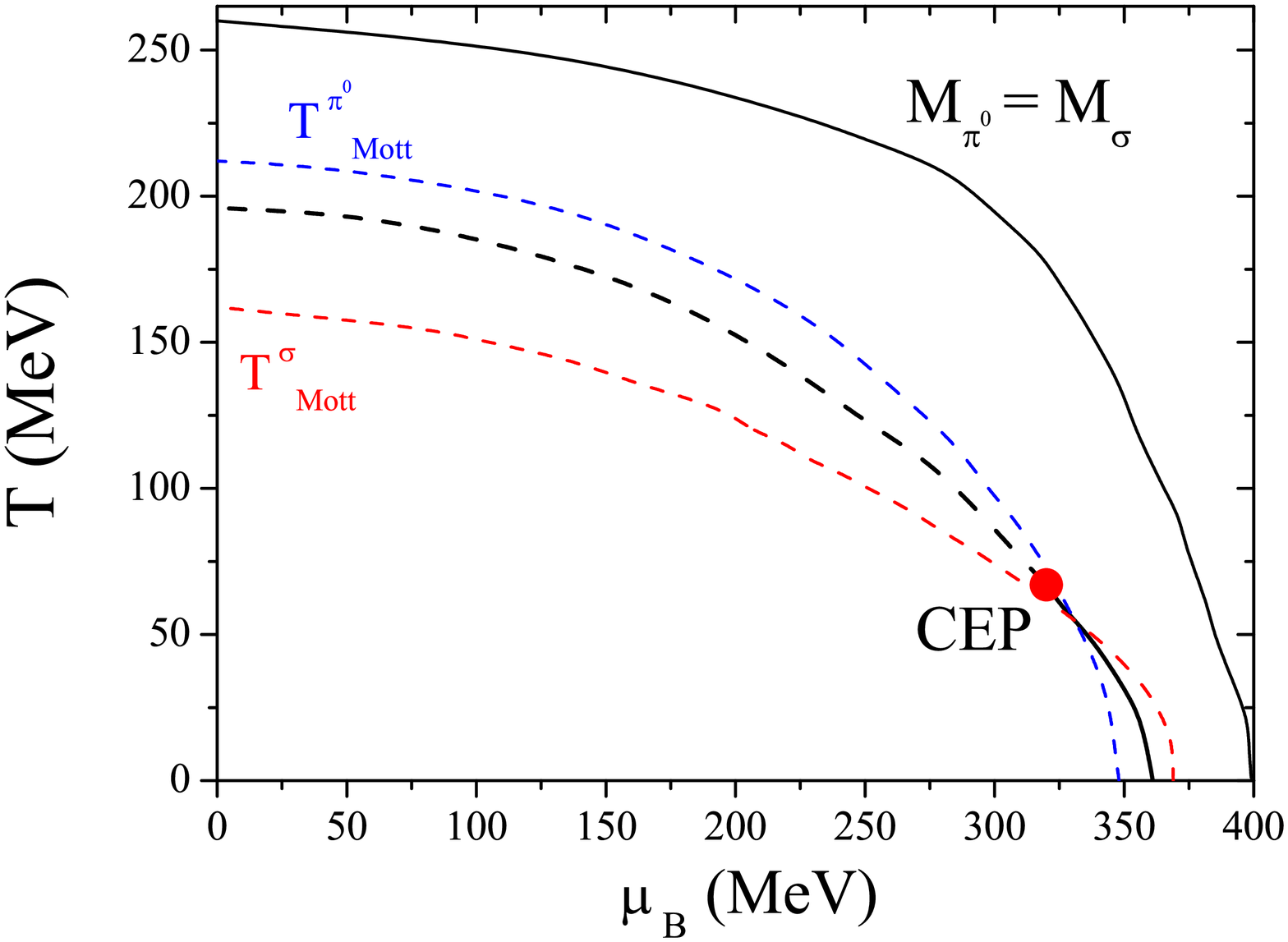,width=8.5cm,height=8cm} \\
  \end{tabular}
\end{center}
\vspace{-1.0cm}\caption{ The \textit{effective} restoration of chiral symmetry, the phase
transition and   the Mott lines for $\pi^{0}$ and $\sigma$ mesons in the
$T-\rho_B(\mu_B)$ plane.}
\label{Fig:7}
\end{figure}

Interesting information can be obtained by calculating the masses of the $\pi^0$ and
$\sigma$ mesons as a function of $T$ and $\rho_B (\mu_B)$ which allows us to obtain a curve
in the $T-\rho_{B}(\mu_B)$ plane. This curve defines the line where the mesons became
degenerate (Fig. \ref{Fig:7}). In Fig. \ref{Fig:7} we also
represent the \textquotedblleft Mott lines\textquotedblright  for the $\pi^{0}$ and  the
$\sigma$, as well as the critical line. As can be seen, the  phase transition
associated to \textit{partial} restoration of chiral symmetry occurs above the Mott line
for the pion and below the Mott line for the sigma, in most of the first order phase
transition region;  the opposite happens  in the crossover region. Concerning the
\textit{effective} restoration of chiral symmetry, one can see, from the line of
convergence of the chiral partners, that it happens after the \textit{partial}
restoration of chiral symmetry and the dissociation of the two mesons.

As we already saw,  there are dramatic changes in the behavior of
some thermodynamic functions such as the specific heat and the quark number
susceptibilities around the CEP.
So, due to their role as signals for restoration of chiral symmetry it is demanding   to
discuss the behavior of the chiral partners ($\pi^0$, $\sigma$).

First we notice, in Fig. \ref{Fig:7}, that the
two Mott lines cross in the first order region at a point just bellow  the CEP.  This is
probably a remnant of the situation in the chiral limit where the transition is second
order and the pion and sigma dissociate  at the same point.

\begin{figure}[t]
\begin{center}
    \epsfig{file=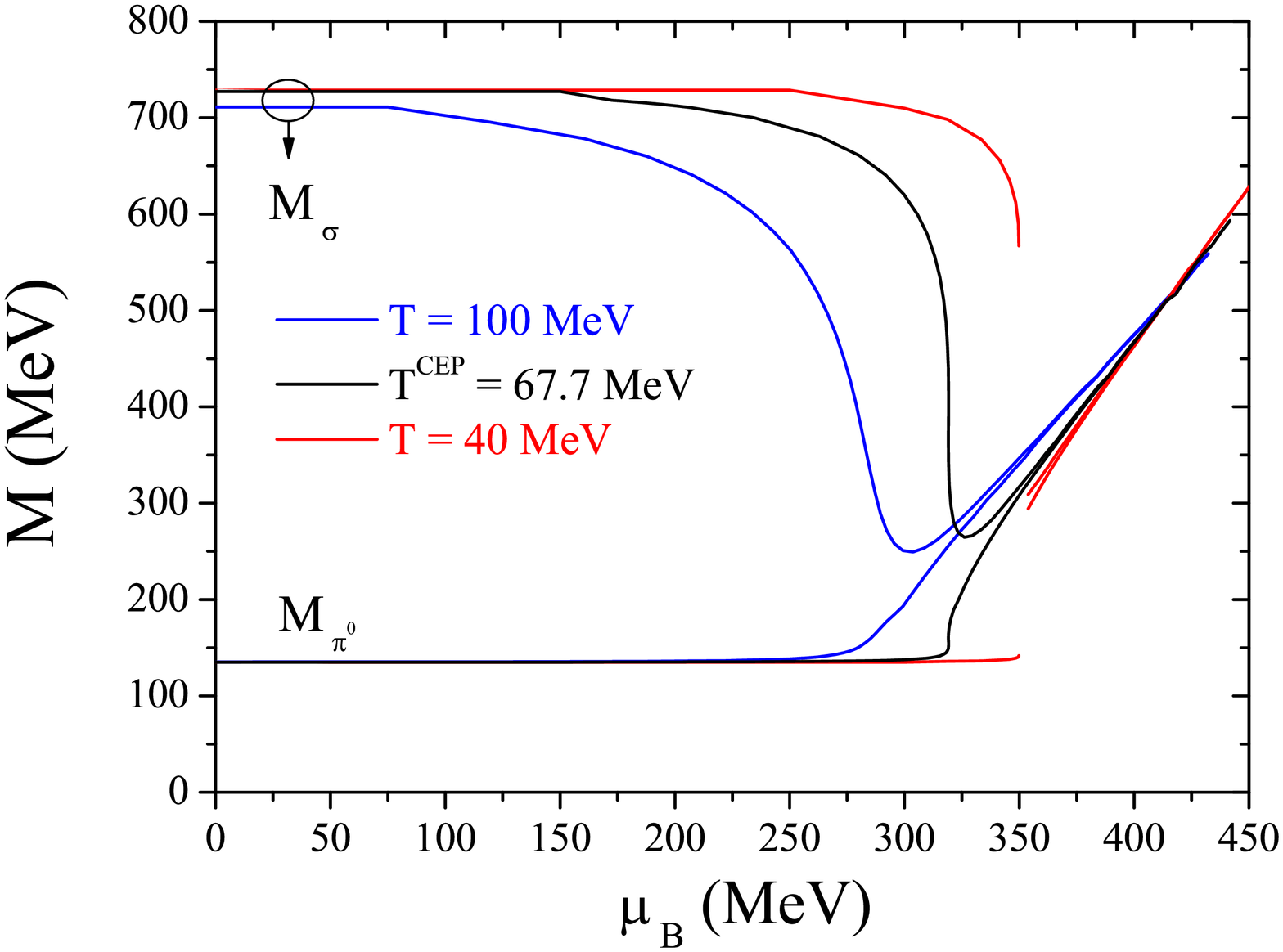,width=10cm,height=9cm}
\end{center}
\vspace{-1.0cm} \caption{Degeneracy of the chiral partners ($\pi^0$, $\sigma$) for
different temperatures around the CEP.}
\label{Fig:8}
\end{figure}

In Fig. \ref{Fig:8} we plot the pion and sigma masses as functions of the
baryonic chemical potential for three different temperatures: $T=40$ MeV $<T^{CEP}$,
$T^{CEP}=67.7$ MeV and $T=100$ MeV $>T^{CEP}$. For $T=40$ MeV and $\mu_B\approx 350$
MeV, a discontinuity is visible in the evolution of the masses, signaling a first
order phase transition. However, according to our criterion, the \textit{effective}
restoration of the chiral symmetry only happens at
$\mu_B\approx 380$ MeV. At the CEP ($T=67.7$ MeV; $\mu_B =  318.4$ MeV), the sharp
decrease (increase) of the sigma (pion) meson masses reflect the nature of the second
order phase transition. Once again the \textit{effective} restoration of chiral symmetry
only happens at $\mu_B\approx 370$ MeV. When $T=100$ MeV $>T^{CEP}$, we have a crossover
and the meson masses have a smooth behavior. In this case, the \textit{effective}
restoration of the chiral symmetry happens at $\mu_B\approx 355$ MeV.

\section{Conclusions}

The properties of the  QCD transition at vanishing chemical potential depend on the
number of quark flavors and on their masses. The critical temperatures of $T_c\approx 155$ MeV
or as high as $T_c\approx 260$ MeV have been reported in the literature. Presently,
considering also nonvanishing chemical potential, some lattice calculations locate the CEP at
$T\approx 160$ MeV and $\mu_B \approx 360$ MeV \cite{Fodor:2004JHEP}. However,
the existence and location of the CEP are not conclusive even for lattice calculations
\cite{Forcrand}.

We proved that our model calculation has been able to reproduce the qualitative phase
structure features, and we also obtain the location of the CEP. We have obtained, at zero
baryon chemical potential in the SU(3) NJL model, values for  the critical temperature
around $120-200$ MeV.  The transition is first order in the chiral limit
($m_u=m_d=m_s=0$). Furthermore, when $m_u=m_d=0$ and $m_s>m_{s}^{crit}$
($m_{s}^{crit}=18.3$ MeV) the transition is of second order ending in a first order line
at the TCP. Finally, when also $m_u=m_d\neq0$, there is a crossover for all  values of
$m_s$ and the location of the CEP depends strongly on the strange quark mass.
Contrarily to what happens in the three-flavor NJL model, we find a
TCP in the two-flavor NJL model in the chiral limit.
This agrees with what  is expected at $\mu_B=0$: for $m_i=0$  the chiral restoration
happens via a second order phase transition for $N_f = 2$, and via a first order for
$N_f\geq3$.
For realistic  values  of the current quark masses the CEP is
located at $T^{CEP} = 79.9$ MeV and $\mu^{CEP}_B = 331.7$ MeV for
$N_f=2$, and at $T^{CEP} = 67.7$ MeV and $\mu^{CEP}_B = 318.4$ MeV
for $N_f=3$.

The pattern characteristic of a first order phase transition has also been analyzed 
through several equations of state and the latent heat. For example, we verified 
that states (droplets) in mechanical equilibrium with the vacuum state at $P=0$ 
are found at zero temperature. 
This leads to  nontrivial consequences for the behavior of the isentropic 
trajectories which terminate at $T=0$ at the first order transition line. 
Our convenient choice of the model parameters, which allows for a first order 
phase transition that is stronger than in other treatments of the NJL model, 
is crucial  to attain this result.

We have studied the baryon number susceptibility and the specific heat which are related
with event-by-event fluctuations of $\mu_B$ or $T$ in heavy-ion collisions. In the two
and three-flavor NJL models, for $\chi_B$, we conclude that the obtained critical
exponents around the CEP in both models are consistent with the mean field values
$\epsilon=\epsilon'=2/3$.
For the specific heat  we obtain nontrivial exponents $1/2<\alpha<2/3$ around the CEP,
indicating a crossover of different universality classes
\cite{Hatta:2003PRD,Schaefer:2006}. This effect is more clearly visible for the critical
exponent of the specific heat  in the SU(2) version of the NJL model, where a crossover
from $\alpha$ to $\alpha_1$ is also observed.
Nevertheless we notice that the values of $\alpha$ in the TCP and in the CEP are
consistent within both versions of the NJL model.
A better insight to the difficult task of the analysis of the phase diagram of QCD can
be provided by an extension of the NJL model where quarks interact with the temporal
gluon field represented by the Polyakov loop dynamics
\cite{Ratti,Hubert,Sasaki,Megias:2006PRD}. Work in this direction is in progress.

Concerning the behavior of the chiral partners in the vicinity  of the CEP, we verified
that the two Mott lines, respectively, for $\sigma$ and $\pi^0$ cross at a point just
bellow  the CEP. On the other hand, there is a sharp decrease (increase) of the sigma
(pion) meson masses at the CEP, which reflects the nature of the second order phase
transition at this point.

\begin{acknowledgments}
Work supported by Grant No. SFRH/BPD/23252/2005 (P. Costa), Centro de
F\'{\i}sica Te\'orica, FCT under Project No. POCI/FP/63945/2005 and
under Project No. PTDC/FP/63912/2005 (GTAE).
\end{acknowledgments}


\appendix
\section{}

In this appendix we give the integrals appearing in the meson propagators, in the vacuum
and at finite temperature and density, as well as some useful expressions concerning
the $\sigma$ meson.

\vspace{0.5cm}

The integrals $I_{1}^{i}$ and $I_{2}^{ij}(P_{0})$ are given by
\begin{align}
\label{firstt}I_{1}^{i} (T\,,\mu_{i} )= \frac{N_{c}}{4\pi^{2}} \int
\frac{\mathtt{p}^{2} d\mathtt{p}}{E_{i}} \left(1- n_{i} - \bar{n}_{i}\right),
\end{align}
\begin{align}
I_{2}^{ii}(P_{0}, T, \mu_{i}) =  \frac{N_{c}}{2\pi^{2}}
{\mathcal{P}} \int\frac{\mathtt{p}^{2} d \mathtt{p}}{E_{i}} \,\, \frac
{1}{4 E_{i}^{2}-P_{0}^{2}} \left(1- n_{i} - \bar{n}_{i}\right),
\end{align}
where $E_{i}=\sqrt{\mathtt{p}^{2}+M_{i}^{2}}$ is the quark energy. To
regularize the integrals we introduce the 3-dimensional cutoff parameter
$\Lambda$. When $P_{0}>2M_{i}$ it is necessary to take into account the
imaginary part of the second integral. It may be found, with help of the
$i\epsilon$ -prescription $P_{0}^{2}\rightarrow P_{0}^{2}-i\epsilon$. Using
\begin{equation}
\lim_{\epsilon\rightarrow0^{+}}\frac{1}{y-i\epsilon}={\mathcal{P}}\frac{1}%
{y}+i\pi\delta(y)
\end{equation}
we obtain the integral
\begin{align}
I_{2}^{ii}(P_{0}, T, \mu_{i}) =  &  \frac{N_{c}}{2\pi^{2}}
{\mathcal{P}} \int\frac{\mathtt{p}^{2} d \mathtt{p}}{E_{i}} \,\, \frac
{1}{4 E_{i}^{2}-P_{0}^{2}} \left(1- n_{i} - \bar{n}_{i}\right) \nonumber\\
&  + i \frac{N_{c}}{16\pi} \sqrt{ 1- \frac{4 M_{i}^{2}}{P_{0}^{2}} } \left[
1-n_{i}\left(\frac{P_{0}}{2}\right) - \bar{n}_{i} \left(\frac{P_{0}}{2}\right)\right].
\end{align}

\vspace{0.5cm}

Concerning the calculation of the propagator for the $\sigma$ meson, the projector
${S}_{ab}$ and the polarization operator ${\Pi}_{ab}^{S}$ matrices, in the case
$\left\langle\bar{q}_{u}\,q_{u}\right\rangle=\left\langle\bar{q}_{d}\,q_{d}\right\rangle$,
have the nonvanishing elements
\begin{align}
S_{33}  &  =g_{S}-g_{D}\left\langle\bar{q}_{s}\,q_{s}\right\rangle,\\
S_{00}  &  =g_{S}+\frac{2}{3}g_{D}\left(  \left\langle\bar{q}_{u}\,q_{u}\right\rangle
+\left\langle\bar{q}_{d}\,q_{d}\right\rangle+\left\langle\bar{q}_{s}\,q_{s}\right\rangle\right)  ,\\
S_{88}  &  =g_{S}-\frac{1}{3}g_{D}\left(  2\left\langle\bar{q}_{u}\,q_{u}\right\rangle
+2\left\langle\bar{q}_{d}\,q_{d}\right\rangle-\left\langle\bar{q}_{s}\,q_{s}\right\rangle\right)  ,\\
S_{08}  &  =S_{80}=-\frac{1}{3\sqrt{2}}g_{D}\left(\left\langle\bar{q}_{u}\,q_{u}\right\rangle
+\left\langle\bar{q}_{d}\,q_{d}\right\rangle-2\left\langle\bar{q}_{s}\,q_{s}\right\rangle\right)  .
\end{align}
Analogously, we get
\begin{align}
\Pi_{00}^{S}(P_{0})  &  =\frac{2}{3}\left[  J_{uu}^{S}(P_{0})+J_{dd}^{S}(P_{0})
+J_{ss}^{S}(P_{0})\right]  ,\\
\Pi_{88}^{S}(P_{0})  &  =\frac{1}{3}\left[  J_{uu}^{S}(P_{0})+J_{dd}^{S}(P_{0})
+4J_{ss}^{S}(P_{0})\right]  ,\\
\Pi_{08}^{S}(P_{0})  &  =\Pi_{80}^{S}(P_{0})=\frac{\sqrt{2}}{3}\left[J_{uu}^{S}(P_{0})
+J_{dd}^{S}(P_{0})-2J_{ss}^{S}(P_{0})\right]  ,
\end{align}
where
\begin{equation}
J_{ii}^{S}(P_{0})=4[2I_{1}^{i}+[P_{0}^{2}-4M_{i}^{2}]I_{2}^{ii}(P_{0})].
\end{equation}
%

\end{document}